\documentclass[onecolumn, fleqn,usenatbib]{mnras}

\usepackage{newtxtext,newtxmath}
\usepackage[T1]{fontenc}

\def\apjl{ApJ}%% Astrophysical Journal, Letters
\def\apjs{ApJS}%% Astrophysical Journal, Supplement
\def\ao{Appl.~Opt.}%% Applied Optics
\def\aap{A\&A}%% Astronomy and Astrophysic
\newcommand{\be}{\begin{equation}}
\newcommand{\ee}{\end{equation}}
\newcommand{\bary}{\begin{eqnarray}}
\newcommand{\eary}{\end{eqnarray}}

\DeclareRobustCommand{\VAN}[3]{#2}
\let\VANthebibliography\thebibliography
\def\thebibliography{\DeclareRobustCommand{\VAN}[3]{##3}\VANthebibliography}

\usepackage{epstopdf}
\usepackage{ulem}

\usepackage{amssymb}
\usepackage{times}
\usepackage{graphicx}
\usepackage[usenames,dvipsnames]{pstricks}
\usepackage{psfrag}
\usepackage{lipsum}
\usepackage{natbib}
\usepackage{textcase}
\usepackage{orcidlink}

\usepackage{multirow}
\usepackage{graphicx}

\usepackage[lofdepth,lotdepth]{subfig} %%%%Package added by 
\title[GRB~250704B/EP250704a]%{Magnetar as progenitor of GRB~250704B}
{GRB~250704B/EP250704a a Short Gamma-Ray Burst Powered by a Magnetar}

\author[Fraija, N. et al.]{
Nissim Fraija\orcidlink{0000-0002-0173-6453},$^{1}$\thanks{E-mail: nifraija@astro.unam.mx}
Antonio Galv\'an\orcidlink{0000-0001-5193-3693},$^{2}$
Boris Betancourt Kamenetskaia\orcidlink{0000-0002-2516-5739},$^{3}$
and Maria G. Dainotti\orcidlink{0000-0003-4442-8546}$^{4,5}$
\\
$^{1}$Instituto de Astronom\' ia, Universidad Nacional Aut\'onoma de M\'exico, Circuito Exterior, C.U., A. Postal 70-264, 04510 M\'exico City, M\'exico\\
$^{2}$Instituto de F\'sica, Universidad Nacional Aut\'onoma de M\'exico, Circuito Exterior, C.U., A. Postal 70-264, 04510 M\'exico City, M\'exico\\
$^{3}$Cosmology, Gravity, and Astroparticle Physics Group, Center for Theoretical Physics of the Universe,
Institute for Basic Science (IBS), Daejeon, 34126, Korea\\
$^{4}$Division of Science, National Astronomical Observatory of Japan, 2-21-1 Osawa, Mitaka, Tokyo 181-8588, Japan \\
$^{5}$The Graduate University for Advanced Studies (SOKENDAI), Shonankokusaimura, Hayama, Miura District, Kanagawa 240-0115, Japan
}

\date{Accepted XXX. Received YYY; in original form ZZZ}

\pubyear{\the\year{}}

\begin{document}
\label{firstpage}
\pagerange{\pageref{firstpage}--\pageref{lastpage}}
\maketitle

\begin{abstract}

GRB~250704B/EP250704a, identified as a short gamma-ray burst (sGRB), exhibited prolonged X-ray emission following the prompt phase. The event also showed an unusual optical and infrared (IR) plateau lasting nearly one day, followed by a rapid achromatic decline. This event was extensively monitored by multiple space- and ground-based observatories across the electromagnetic spectrum. We present temporal and spectral analyses of the prompt, extended X-ray, and afterglow emission spanning radio to gamma-ray energies during the first two days after the burst. The observed multi-wavelength properties are consistent with a long-lived accreting millisecond magnetar remnant. In this framework, the extended X-ray emission is interpreted as internal dissipation within a relativistic magnetized outflow powered by the magnetar spin-down luminosity, whereas the optical/IR and radio emission are explained as synchrotron forward-shock radiation with continuous energy injection. The steep late-time decay is explained through temporal evolution of the microphysical parameters during the post-jet-break phase. The X-ray observations can be interpreted as a superposition of internal dissipation and synchrotron forward-shock emission. These results favour the presence of a long-lived magnetar remnant powering the multi-wavelength evolution of GRB~250704B/EP250704a.

\end{abstract}


%%%%%%%%%%%%%%%%%%%%%%%%%%%%%%%%%%%%%%%%%%%%%%%%%%

%%%%%%%%%%%%%%%%% BODY OF PAPER %%%%%%%%%%%%%%%%%%

\section{Introduction}

Short gamma-ray bursts (sGRBs) are widely believed to originate from mergers of compact-object binaries involving neutron stars (NSs) and/or black holes (BHs) \citep{1992ApJ...395L..83N, 1992ApJ...392L...9D, 1992Natur.357..472U, 1994MNRAS.270..480T, 2011MNRAS.413.2031M}.  The prevailing scenario for the compact-object remnant involves either a rapidly accreting BH \citep[e.g., see][]{2000ApJ...528L..29B, 2006PhRvD..73f4027S, 2011ApJ...732L...6R} or a fast-spinning, strongly magnetized, long-lived NS \citep[a millisecond magnetar; e.g.,][]{1992ApJ...392L...9D, 1992Natur.357..472U,  2011MNRAS.413.2031M}.  In the first scenario, the merger of compact objects can produce a BH surrounded by an accretion torus. The energy released through accretion, or through accretion-mediated extraction from a Kerr BH \citep{1977MNRAS.179..433B}, drives a relativistic outflow. In the second scenario, a millisecond magnetar is created with sufficient rotational energy to prevent gravitational collapse \cite[e.g., see][]{2011MNRAS.413.2031M}.  The magnetic dipole spin-down luminosity of the magnetar then powers the relativistic outflow and sustains prolonged central-engine activity.\\ 

Internal shocks \citep{1994ApJ...430L..93R, 1997ApJ...490...92K, 1998MNRAS.296..275D} and magnetic reconnections \citep{2000ApJ...537..810W, 2003ApJ...596.1080V, 2003ApJ...596.1104V}, which efficiently convert kinetic and magnetic energy into radiation, are commonly proposed as mechanisms for the prompt gamma-ray episode, typically characterized by a total duration of less than two seconds \citep[$T_{\rm 90}\lesssim 2\,{\rm s}$;][]{Kouveliotou1993}. In certain sGRBs, a temporarily extended soft X-ray emission has been observed following the prompt episode \cite[e.g., see][]{2005Natur.437..855V, 2009ApJ...696.1871P, 2014ARA&A..52...43B,  2015ApJ...811....4K, 2017ApJ...848L..14G}.  A subsequent episode, referred to as {\textit afterglow}, is observed across a broad spectrum from radio wavelengths to gamma rays. This phenomenon is typically interpreted using the synchrotron afterglow model, which attributes the emission to non-thermal electrons accelerated in the forward shock (FS). The shock is generated when the outflow interacts with the circumburst medium, transferring a fraction of its energy to the surrounding environment \citep{1999ApJ...513..669K, 2002ApJ...568..820G, 1998ApJ...497L..17S}. The shock energy is partitioned between particle acceleration and magnetic-field amplification through the microphysical parameters $\epsilon_{\rm e}$ and $\epsilon_{\rm B}$, which denote the fractions of the post-shock internal energy transferred to relativistic electrons and magnetic fields, respectively. Due to the limited understanding of energy transfer processes among protons, electrons, and magnetic fields in relativistic shocks, it is plausible that the microphysical parameters may evolve over time. Variations in these parameters have been invoked to model X-ray, optical, and radio afterglow observations \citep[e.g., see][]{2003ApJ...597..459Y, 2006A&A...458....7I,2006MNRAS.369..197F, 2006MNRAS.369.2059P, 2005PThPh.114.1317I, 2006MNRAS.370.1946G, 2020ApJ...905..112F}. 

On the other hand, during the afterglow phase, in a significant fraction of bursts, the X-ray and optical light curves have exhibited a ``plateau" phase \citep[e.g., see][]{2006ApJ...642..354Z, 2006ApJ...642..389N}. Based on its origin, there are two kinds of plateaus: internal or external. Radiation from the magnetar's internal energy dissipation during spin-down explains plateaus of internal origin.  Internal plateaus are generally interpreted as direct signatures of prolonged central-engine activity rather than standard external-shock emission \citep{2007ApJ...665..599T, 2013MNRAS.430.1061R, lu14, 2018ApJ...857...95M}. Plateaus of external origin have been interpreted as refreshed shocks powered by continuous energy injection from either a millisecond magnetar or fallback accretion onto a BH \citep{zhang2001, dallosso2011, 2018ApJ...857...95M, 2019ApJ...872..118B, 2019ApJ...887..254B}

The advent of new wide-field, high-cadence transient missions has greatly improved our ability to detect and localize faint, fast transients. In particular, the combined capabilities of the {\itshape Einstein Probe (EP)}, the {\itshape Space-based multiband astronomical Variable Objects Monitor (SVOM)}, and the {\itshape Neil Gehrels Swift Observatory}  enable the prompt detection, localization, and multi-wavelength characterization of faint and rapidly evolving transients that were previously difficult to capture. This coordinated observational framework is especially powerful for sGRB and GRB-like events, where rapid follow-up is essential for constraining afterglow properties (or their absence). These capabilities open new opportunities to distinguish between multiple physical origins for short-lived high-energy events, including compact binary mergers \citep{Becerra2025,Jonker2026}, off-axis GRBs \citep{Gianfagna2025,Yadav2025,vanDalen2025,Quirola2025}, fast X-ray transients \citep[FXTs;][]{2025ApJ...985...21Z,2025A&A...701A.225B}, and other exotic progenitor channels \citep{OConnor2025}.

On July 4, 2025,  the {\itshape EP} Wide-field X-ray Telescope \citep[WXT;][]{2025GCN.40941....1L} and the Gamma-Ray Monitor (GRM) instrument onboard {\itshape SVOM} detected a short-duration burst designated EP250704a/GRB~250704B \citep{2025GCN.40940....1S}.   This event was subsequently detected across gamma-ray, X-ray, optical/IR, and radio wavelengths by multiple space- and ground-based observatories \citep{2025GCN.41025....1S, 2025GCN.40972....1F, 2025GCN.40951....1E, 2025GCN.40942....1S, 2025GCN.40945....1M, 2025GCN.41038....1S,2025GCN.41046....1R}. Based on the identification of several absorption lines, attributed to the \ion{Mg}{II} doublet, various \ion{Fe}{II} lines, and \ion{Mg}{I}, a redshift of $z = 0.661$ was reported \citep{2025GCN.40966....1A}.

GRB 250704B/EP250704a exhibited extended X-ray emission together with an optical plateau lasting nearly one day, followed by a late abrupt decay. This manuscript presents a multi-wavelength analysis of GRB~250704B/EP250704a.  The long-lived optical plateau and subsequent abrupt decay are interpreted within a synchrotron FS scenario involving continuous energy injection and temporal evolution of the microphysical parameters. The X-ray evolution is naturally explained by the internal dissipation powered by the spin-down luminosity of a millisecond magnetar together with a contribution from the synchrotron FS. The manuscript is organized as follows. In Section~\ref{sec:obs}, we present the multi-wavelength observations and data reduction of GRB~250704B/EP250704a. In Section~\ref{sec:model}, we model, interpret, and discuss the multi-wavelength observations. Finally, in Section~\ref{sec:summary}, we summarize our main results and conclusions. We adopt the convention $Q_{\rm x}=Q/10^{\rm x}$ in cgs units throughout this paper. 

\section{Observations and Data Analysis: GRB~250704B/EP250704a}
\label{sec:obs}

The multi-wavelength analysis presented in this work is based on publicly available observations reported by the Einstein Probe collaboration \citep{2025GCN.40941....1L,2025GCN.40956....1L}, Swift/XRT \citep{2025GCN.40951....1E}, Konus-Wind \citep{2025GCN.40972....1F}, CALET \citep{2025GCN.41025....1S}, and the extensive optical, infrared, and radio follow-up campaigns reported by \citet{2025GCN.40942....1S,2025GCN.40945....1M, 2025GCN.40958....1G,2025GCN.40971....1M,2025GCN.40975....1L, 2025GCN.41038....1S,2025GCN.41060....1S}, and other GCN Circulars. No new observations are presented in this work; instead, we compile these publicly available data and perform a homogeneous temporal and spectral analysis within the framework of our theoretical model.

\subsection{Observations}
\subsubsection{High Energies}

On 2025 July 4 at 08:16:52 UTC, the GRM/{\itshape SVOM} instrument detected a short-duration burst, designated GRB~250704B, temporally and spatially coincident with EP250704a \citep{2025GCN.40940....1S}. The GRM light curve exhibited two distinct spikes, with a measured duration of $T_{90} = 0.68^{+0.16}_{-0.14}$~s in the 15-5000~keV band, consistent with the classification of a short/hard GRB \citep{2025GCN.40940....1S}. Additional detections of this source were reported by the CALorimetric Electron Telescope ({\itshape{CALET}}/Gamma-ray Burst Monitor \citep[GRBM);][]{2025GCN.41025....1S}, {\textit Konus-Wind} \citep{2025GCN.40972....1F}, and {\textit Insight-HXMT}/HE \citep{2025GCN.40978....1W}.\\

Simultaneously, the {\itshape EP}-WXT \citep{2025GCN.40941....1L} triggered the same source, designating it as EP250704a. The automated localization yielded a position of RA = 20:03:29.28, Dec = +12:01:48.00 (J2000) with a 90\% confidence error radius of $\sim$3\arcmin, including systematic uncertainties. A follow-up pointing by the EP Follow-up X-ray Telescope (FXT), commencing $\sim$120~s after trigger, revealed an uncatalogued X-ray source at RA = 20:03:29.81, Dec = +12:01:27.48 with an uncertainty of $\sim20\arcsec$ \citep{2025GCN.40941....1L}. 
The WXT light curve shows a transient lasting approximately $\sim 10\, {\rm s}$, significantly longer than the prompt gamma-ray duration measured by GRM/SVOM. The time-averaged spectrum in the 0.5--4~keV band can be described by an absorbed power law with photon index $\Gamma \approx 1.7$ (with Galactic absorption fixed at $N_{\rm H} = 8\times10^{20}$~cm$^{-2}$), corresponding to an average unabsorbed flux of $\sim$1.3$\times$10$^{-9}$~erg~cm$^{-2}$~s$^{-1}$ \citep{2025GCN.40956....1L}.\\

The {\itshape Swift}/X-ray Telescope (XRT) began an automatic follow-up 2052~s after the trigger \citep{2025GCN.40951....1E}, detecting a fading X-ray counterpart consistent with the optical counterpart reported by \citet{2025GCN.40942....1S}. The {\itshape Swift}/XRT reported a position of RA=20:03:29.18 and Dec=+12:01: 23.9 with an uncertainty of $3.6\arcsec$. 
The XRT instrument monitored GRB 250704B/EP250704a in Photon Counting mode between $3.269\times 10^3$ to $7.831\times 10^3\,{\rm s}$ after trigger. Spectral analysis revealed both Galactic and intrinsic absorption components.\footnote{The values reported of Galactic and intrinsic absorption column density were $N_{\rm H}=1.47\times 10^{21}\,{\rm cm^{-2}}$ and $2.47\times 10^{20}\,{\rm cm^{-2}}$, respectively, for $3269\,{\rm s}$, and  $N_{\rm H}=1.47\times 10^{21}\,{\rm cm^{-2}}$ and $1.89\times 10^{21}\,{\rm cm^{-2}}$, respectively, for $7673\,{\rm s}$, after the trigger time.}

%%%%%%%%%%%%%%%%%%%%%%%%%%%%% 3er keV emission %%%%%%%%%%%%%%%%%%

%\subsubsection{Multi-wavelength Follow-Up}
%The optical counterpart of GRB~250704B/EP250704a was identified by COLIBRÍ \citep{2025GCN.40942....1S}. Extensive optical and IR follow-up observations were carried out by several facilities including Very Large Telescope \citep[VLT;][]{2025GCN.40945....1M}, the Rapid Eye Mount 60 cm robotic Telescope \citep[REM;][]{2025GCN.40947....1B}, Pan-STARRS \citep{2025GCN.40958....1G}, {\itshape SVOM}/VT \citep{2025GCN.40960....1X}, the Nordic Optical Telescope \citep[NOT;][]{2025GCN.40971....1M}, and Las Cumbres Observatory \citep{2025GCN.40975....1L}, among others.

%At radio wavelengths, the burst was monitored by VLA \citep{2025GCN.41038....1S} and MeerKAT \citep{2025GCN.41060....1S}.

%Finally, in radio, the burst was monitored by the Very Large Array (VLA) at 6 and 10~GHz,  \citep{2025GCN.41038....1S,2025GCN.41046....1R} and MeerKAT at 1.3~GHz \citep{2025GCN.41060....1S}.% and uGMRT at 1.3 and 0.65~GHz.

\subsection{Data Analysis}

%\subsubsection{High-Energies}
%The left-hand panel in Figure \ref{fig1:gama_LC} shows
%The standard deviation is the error shown in the right panel of Figure \ref{fig1:gama_LC}.
%\footnote{\url{https://cgbm.calet.jp/cgbm_trigger/ground/1435651911/}}
\subsubsection{Prompt episode and Extended emission}
\begin{figure}
\resizebox*{0.5\textwidth}{0.35\textheight}
{\includegraphics{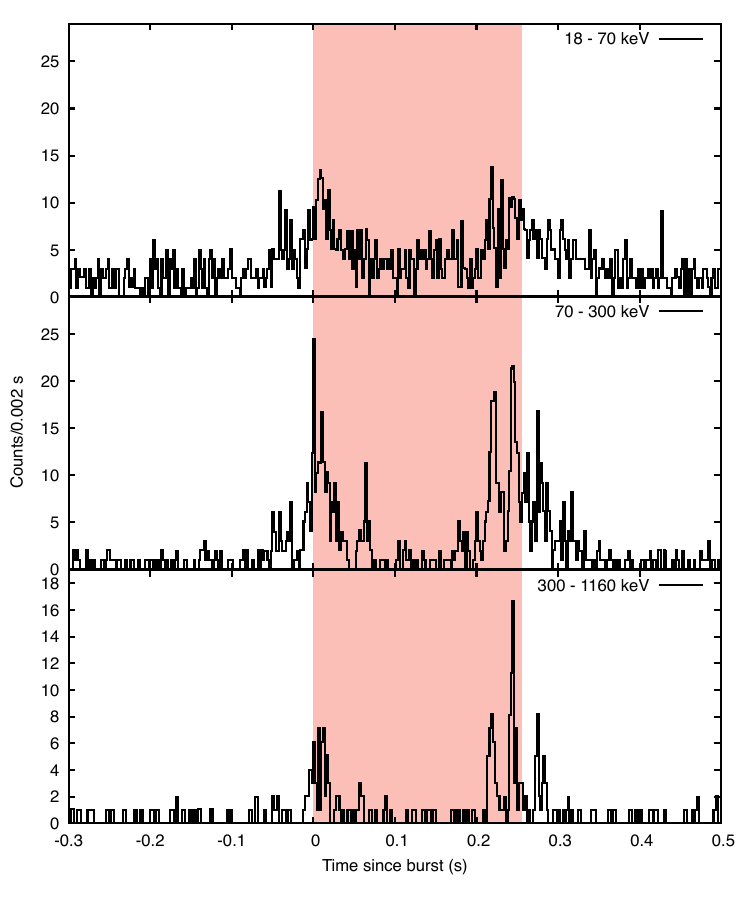}}
\resizebox*{0.5\textwidth}{0.35\textheight}
{\includegraphics{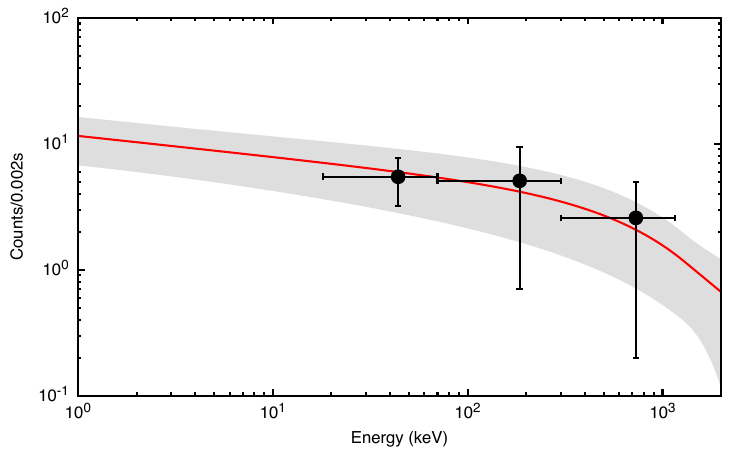}}
{ \centering
\resizebox*{0.55\textwidth}{0.5\textheight}
{\includegraphics{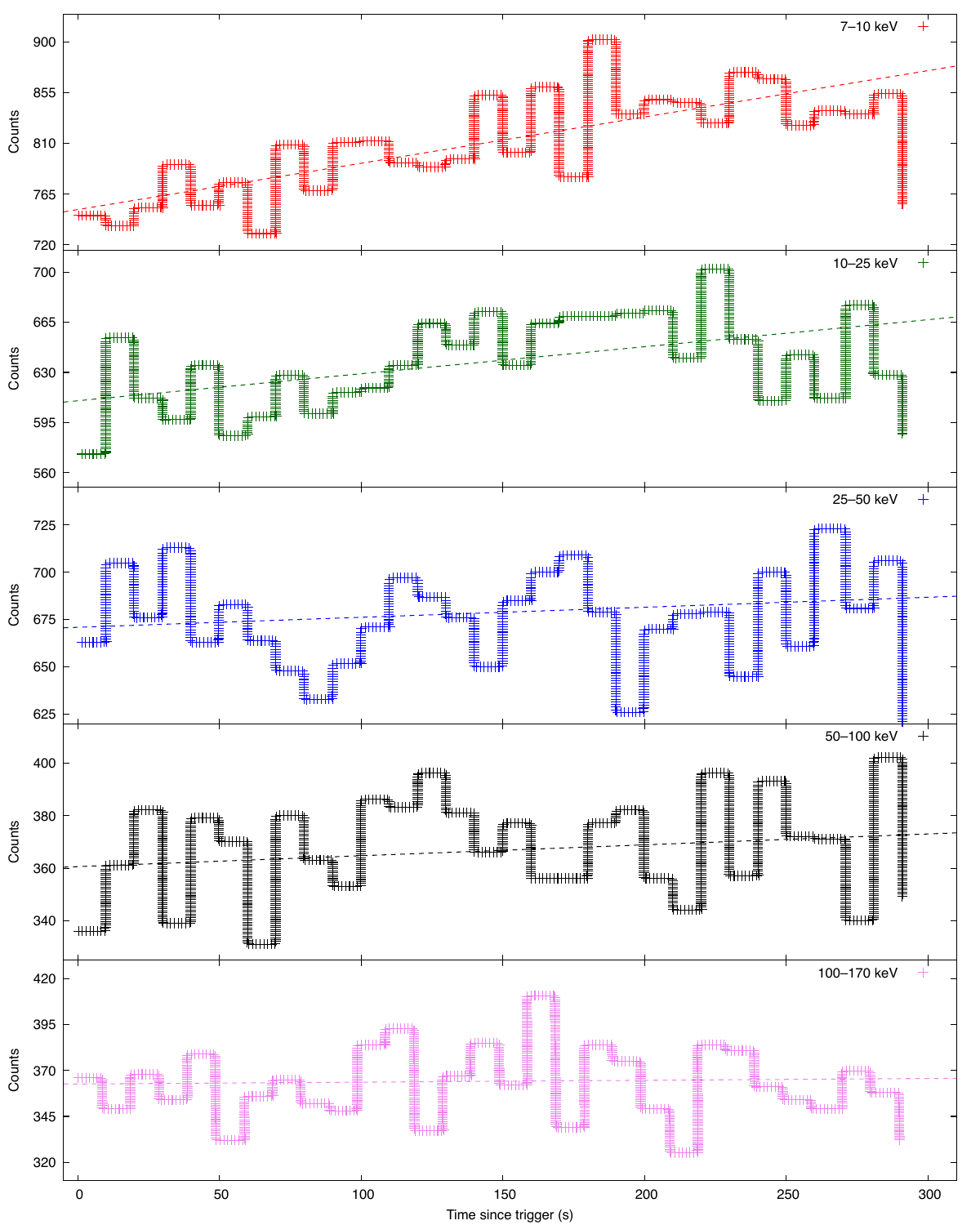}}
%\resizebox*{0.5\textwidth}{0.35\textheight}
%{\includegraphics{Figuras/GRB250907B_Lightcurve.pdf}}
}
\caption{The upper-left panel presents the {\itshape KW} public light curves of GRB~250704B/EP250704a in the 18-70, 70-300, and 300-1160~keV channels, arranged from top to bottom. The orange region indicates the period used for spectral data collection. The upper-right panel displays the corresponding spectrum. The rate values across these three channels are averaged and fit to a Gaussian function. The least squares method fits the spectrum with a Band function.  Bottom panel: The HXM2 ({\itshape CALET}/CGBM) light curve on counts in five channels (from top to bottom): 7 - 10, 10 - 25, 25 - 50, 50 - 100, and  100 - 170~keV. The dotted lines in each channel correspond to the best-fit straight line.}
 \label{fig1:gama_LC}
\end{figure}

The upper panels in Figure \ref{fig1:gama_LC} show the gamma-ray light curve (left) and spectrum (right) of GRB~250704B/EP250704a.  The light curve of the prompt phase exhibited in three subpanels (from top to bottom) $18 - 70$, $70 - 300$ and $300 - 1160~\rm keV$, was  built using the public {\itshape KW} database.\footnote{\url{http://www.ioffe.ru/LEA/GRBs/GRB250704_T29791/kw20250704_29791_2ms.pdf}}  The light curve exhibited two different structured pulses separated $\sim 1.5\,{\rm s}$, with the separation becoming more pronounced at higher energies.
 
%Given the distance of this burst \citep[$z=0.661$;][]{2025GCN.40966....1A}, GRB~250704B aligns with the Type I (short/hard) burst population in the Amati and Yonetoku diagrams.\footnote{\url{http://www.ioffe.ru/LEA/GRBs/GRB250704\_T29791/GRB250704B\_rest\_frame.pdf}} 
The orange region in this panel corresponds to the selected period of data to build the spectrum. We extracted the time-averaged spectrum of GRB~250704B/EP250704a using these three channels averaging the rate values in those channels and fit them with a Band function using the least squares technique \citep{1993ApJ...413..281B}. The best-fit values of the Band function parameters ($E_{\rm p}$, $\alpha_{\gamma}$ and $\beta_{\gamma}$) are listed in Table \ref{tab1:fit}.
Taking into account the distance ($z=0.661$), the peak energy ($E_{\rm p}=970.1\pm 100.6\,{\rm keV}$), low-energy ($\alpha_{\gamma}=-1.160\pm0.086$) and high-energy ($\beta_{\gamma}=-2.28\pm1.13$) power-law (PL) indexes, the isotropic gamma-ray energy in the 18 - 1160~keV band is $E_{\rm \gamma, iso}= (5.28\pm0.76)\times 10^{51}\,{\rm erg}$. The rest-frame peak spectral energy becomes $E_{\rm p, z}=(1611.17\pm 100.6)\,{\rm keV}$. The parameter values of the Band function are 
consistent with previous reports \citep{2025GCN.40972....1F}.\\

The lower panel in Figure \ref{fig1:gama_LC} shows the HXM2 ({\itshape CALET}/CGBM) 10s-binning light curve on counts lasting almost 300 s in five channels (from top to bottom): 7 - 10, 10 - 25, 25 - 50, 50 - 100, and  100 - 170~keV.\footnote{\url{https://cgbm.calet.jp/cgbm\_trigger/flight/1435651911/index.html}}   The public {\itshape CALET} database shows variability ($\Delta t / t \ll 1$) in all channels. The count rate shows an overall increasing trend with time, particularly in the lowest-energy bands. We fit the temporal evolution in each channel with a PL function using the ROOT software package. The resulting slopes are listed in Table~\ref{tab2:fit}. The soft extended emission is most prominent in the 7–10 keV band and persists for approximately $\gtrsim 300$ s, clearly exceeding the duration of the prompt gamma-ray episode ($T_{90}=0.68^{+0.16}_{-0.14}$ s).

%The public {\itshape CALET} database shows 0.125s-binning light curves exhibiting variability ($\Delta t/t \ll 1 $) in all channels.   We note that the counts as a function of time increase  monotonically in each panel, being this behaviour more evident in the lowest energy band (7 - 10~keV). Using the ROOT Software package \citep{1997NIMPA.389...81B},  we fit each channel with a PL function as indicated in this panel, and show through the slope ($m_{\rm j}$ with $j$ from $1$ to $5$ associated to channels from 7 - 10 to 100-170~keV) this trend.  We can see in Table \ref{tab2:fit}  that during first $\sim 300\,{\rm s}$  the slope is almost constant ($m_5=0.01\pm0.04$) in the 100 - 170~keV channel and is larger ($m_1=0.41\pm0.06$) in the 7 - 10~keV.  The data reveal an extended emission that is most prominent in the lowest energy band and persists for approximately $\gtrsim 300\,{\rm s}$. This emission is distinct from the gamma-ray prompt episode, which lasts $T_{\rm 90}=0.68^{+0.16}_{-0.14}\,{\rm s}$ \citep{2025GCN.40940....1S}.\\

\subsubsection{Multiwavelength afterglow episode}

\begin{figure}
{ \centering
\resizebox*{0.5\textwidth}{0.35\textheight}
{\includegraphics{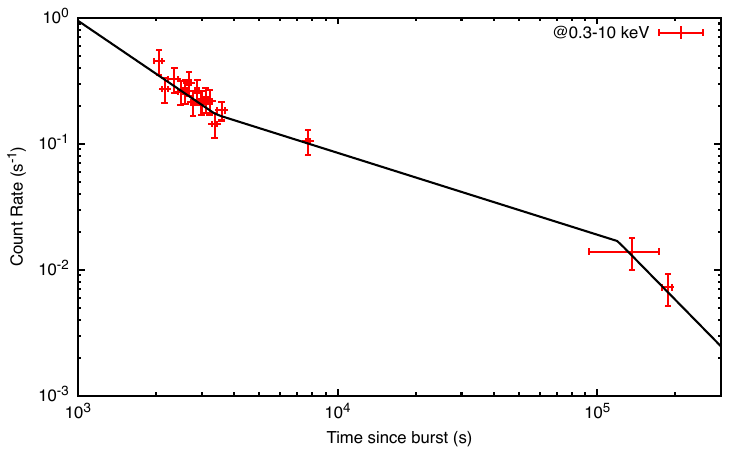}}
\resizebox*{0.5\textwidth}{0.35\textheight}
{\includegraphics{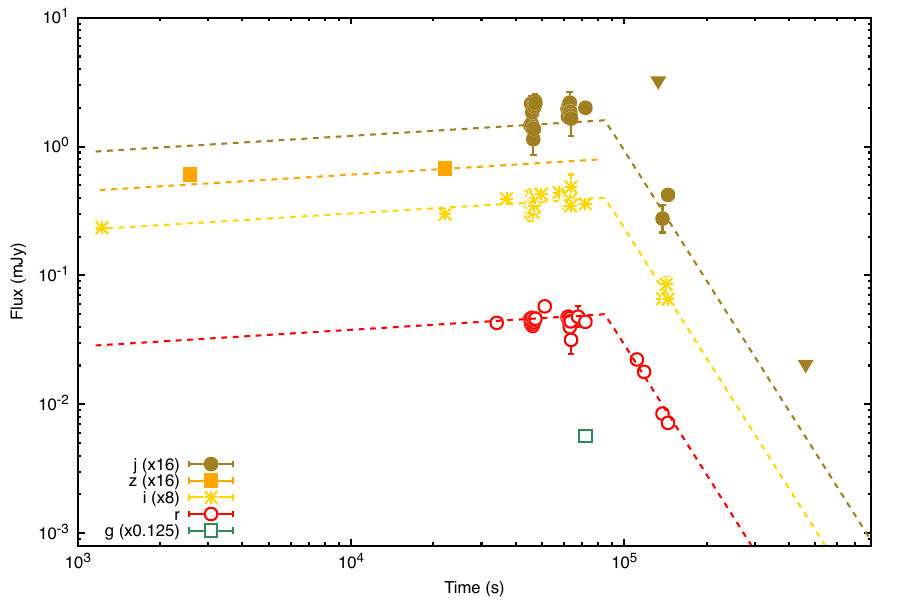}}
\resizebox*{0.5\textwidth}{0.35\textheight}
{\includegraphics{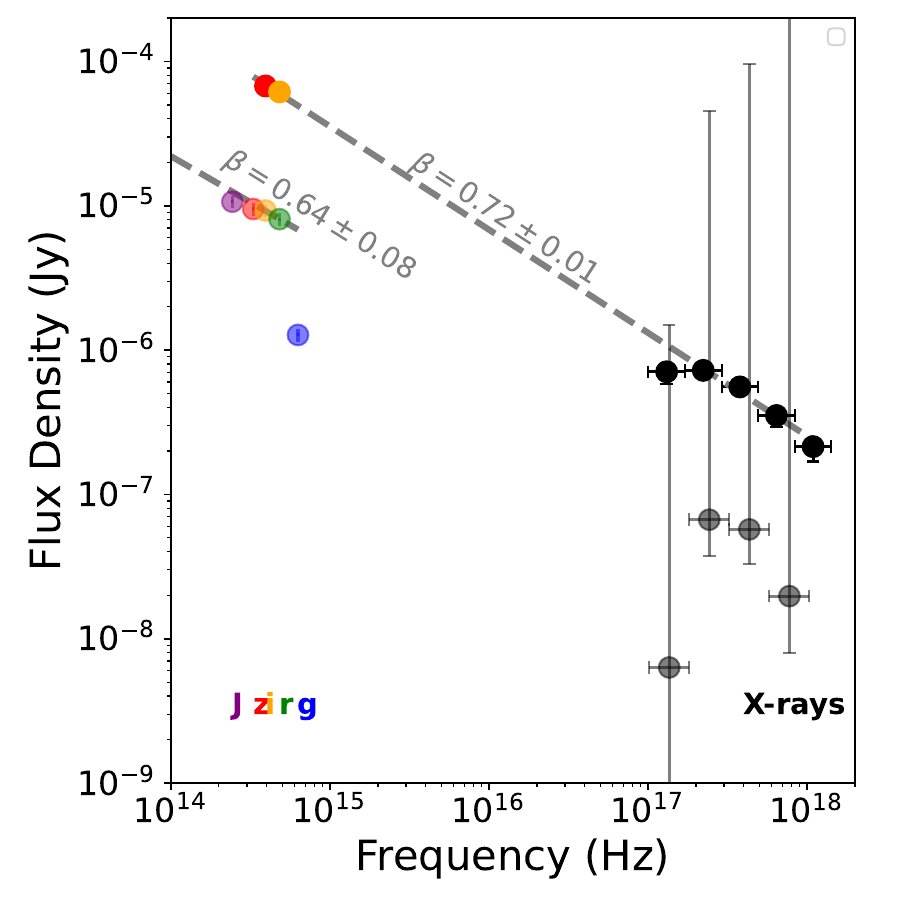}}
}
\caption{The upper panels show the X-ray light curve at 0.3-10~keV (left) and optical/IR light curves (right) in the J, z, i, r and g filters. The lines in both panels correspond to the best-fit PL segments. The lower panel displays the SED of EP250704a at $2.7\times 10^3\, {\rm s}$ and $ 1.5\times 10^3\, {\rm s}$ (lighter points). We combined binned X-ray data from the \textit{Swift}/XRT repository and interpolated the optical flux densities corrected for Galactic extinction.}
 \label{fig3:x-ray_and_optical}
\end{figure}

The upper panels in Figure \ref{fig3:x-ray_and_optical} show the X-ray (left) and optical/IR (right) light curves with the best-fit PL segments ($\propto t^{-\alpha_{\rm i}}$) using the ROOT Software package \citep{1997NIMPA.389...81B}. The X-ray light curve shown at 0.3 - 10~keV was retrieved from the public online repository \citep{2010A&A...519A.102E} hosted by the UK \textit{Swift} Science Data Centre.\footnote{\url{https://www.swift.ac.uk/burst\_analyser/00019908/}}  The optical/IR light curves shown in the {\rm g}, {\rm r}, {\rm i}, {\rm z} and {\rm J} filters are taken from \cite{2025GCN.40942....1S, 2025GCN.40945....1M, 2025GCN.40956....1L, 2025GCN.40958....1G,
2025GCN.40962....1M, 2025GCN.40963....1M,
2025GCN.40965....1L, 2025GCN.40966....1A,
2025GCN.40970....1Y, 2025GCN.40971....1M,
2025GCN.40975....1L, 2025GCN.41024....1V,
2025GCN.41030....1S, 2025arXiv250902769S}.\\   

%%%%%%aqui voy aqui voy aqui voy aqui voy aqui voy aqui voy

For descriptive purposes, we identify three temporal intervals in the X-ray light curve, labeled as ``I'', ``II'' and ``III'' corresponding to the time intervals $\leq 3.4^{+36.9}_{-1.3}\times 10^3\,{\rm s}$, [$3.4^{+36.9}_{-1.3}\times 10^3$ : $1.2^{+0.9}_{-1.2}\times 10^5\,{\rm s}$] and $>1.2^{+0.9}_{-1.2}\times 10^5\,{\rm s}$, respectively. The normal decay in epoch I is characterised by a temporal index of $\alpha_{\rm x,I}=1.4^{+3.6}_{-0.6}$, whereas the subsequent gradual decline observed in epoch II is represented by a temporal index of $\alpha_{\rm x,II}=0.65^{+0.18}_{-0.61}$. In period III, the late steeper decay is characterised with the temporal index $\alpha_{\rm x,III}=2.1^{+5.9}_{-1.9}$.  The spectral analyses of X-rays performed by the Swift-XRT team\footnote{\url{https://www.swift.ac.uk/xrt\_spectra/00019908/}} led to the best-fit values of $\beta_{\rm x}=\Gamma_{\rm x}-1=0.96^{+0.16}_{-0.13}$ and $0.93^{+1.02}_{-0.49}$, at $3.2\times 10^3$ and $7.7\times 10^3\,{\rm s}$, respectively.

The best-fit values of the temporal indexes ($\alpha_{\rm i}$) and temporal breaks as reported by the \textit{Swift}/XRT Team at the \textit{Swift} page\footnote{\url{https://www.swift.ac.uk/xrt\_live_cat/00019908/}} are listed in Table \ref{tab3:fit}.  We identify two intervals in the optical/IR light curves separated by the temporal break at $\sim 7.7\times 10^{4}\,{\rm s}$; a plateau phase described with temporal indexes $ -0.13\lesssim \alpha_{\rm o, I}\lesssim 0.10$, followed by a steep decay with indexes $ 3.29\lesssim \alpha_{\rm o, II}\lesssim 3.35$.   The best-fit values found of temporal indexes and breaks from each optical/IR filter are listed in Table \ref{tab4:fit}. The lower panel in Figure \ref{fig3:x-ray_and_optical} shows the Spectral Energy Distributions (SEDs) of EP250704a at $2.7\times 10^3$ and $1.5\times 10^5\, {\rm s}$ with the best-fit line. The spectral indexes obtained at these epochs from the best-fit model are $\beta_{\rm o}=\beta_{\rm x} = 0.72 \pm 0.01$ and $0.64 \pm 0.08$ at $2.7\times 10^3$ and $1.5\times 10^5\, {\rm s}$, respectively. We combine binned X-ray data from the \textit{Swift}/XRT repository and interpolated the optical flux densities (corrected for Galactic extinction).   The spectral index obtained at $2.7\times 10^3\,{\rm s}$ is consistent  with the value reported by the XRT-Swift team\footnote{\url{https://www.swift.ac.uk/xrt\_spectra/00019908/}} $\beta_{\rm x}=\Gamma_{\rm x}-1=0.96^{+0.16}_{-0.13}$ at $3.2\times 10^3\,{\rm s}$ using only the X-ray observations in the 0.3 - 10 keV energy range. Because the X-ray data at $1.5\times 10^5\, {\rm s}$ were of insufficient statistical quality, they were excluded from the fitting procedure.

The multi-wavelength observations exhibit three main features: (i) a short prompt gamma-ray episode followed by extended soft X-ray emission, (ii) a long-lasting optical/IR plateau phase, and (iii) a late achromatic steep decay. In the following section, we explore whether these features can be interpreted within a long-lived magnetar scenario involving internal dissipation and synchrotron FS emission with continuous energy injection and evolution of microphysical parameters.

\section{Modelling GRB~250704B/EP250704a: Continuous energy injection into the afterglow from a millisecond magnetar}
\label{sec:model}

%We consider potential candidates 

\subsection{Spin-down millisecond magnetar with accretion}\label{model}

In the magnetar scenario, the rotational energy constitutes the dominant global energy reservoir, while the magnetic energy stored in the magnetosphere governs how this energy is extracted and dissipated \citep{1992ApJ...392L...9D, 1992Natur.357..472U, 1994MNRAS.270..480T, 2011MNRAS.413.2031M}.  The interplay between fallback accretion, magnetic spin-down, and relativistic outflow dissipation determines the observed multi-wavelength evolution
 \citep{2011MNRAS.413.2031M, 2018ApJ...857...95M, 2011ApJ...736..108P}. During the prompt phase, the relativistic outflow is expected to be highly magnetized, allowing magnetic reconnection and internal dissipation processes to efficiently convert Poynting-flux energy into gamma-ray radiation 
 \citep{2003astro.ph.12347L, 2002A&A...387..714D, 2007A&A...469....1G, 2011ApJ...726...90Z}.  At later times, the long-lasting X-ray emission is interpreted as internal dissipation within a relativistic magnetized outflow powered by the magnetar spin-down luminosity, while the magnetization parameter regulates the dissipation regime and radiative efficiency \citep{2011MNRAS.413.2031M, 2018ApJ...857...95M,  Bucciantini2011Magnetars, 2017MNRAS.472.3058B}

%In the magnetar scenario, the rotational energy constitutes the dominant global energy reservoir, while the magnetic energy stored in the magnetosphere governs how this energy is extracted and dissipated. The interplay between fallback accretion, magnetic spin-down, and relativistic outflow dissipation determines the observed multi-wavelength evolution.  During the prompt phase, the relativistic outflow is expected to be highly magnetized, so magnetic reconnection and internal dissipation processes can efficiently convert Poynting-flux energy into gamma-ray radiation.  At later times, the long-lasting X-ray emission is interpreted as internal dissipation within a relativistic magnetized outflow powered by the magnetar spin-down luminosity, while the magnetization parameter regulates the dissipation regime and efficiency.

\subsubsection{Light curve from the Internal Dissipation}

The total available energy reservoir of a millisecond magnetar is its rotational energy, which can be as high as 
\be\label{Erot}
E_{\rm}=\frac12 I\, \Omega^2\,\approx 2.6 \times 10^{52}\,{\rm erg}\,M^{\frac32}_{\rm ns,1.4}\,P^{-2}_{-3}\,,
\ee
where $I$ is the moment of inertia,\footnote{The moment of inertia is $I\simeq 1.3\times 10^{45}\,M^{\frac32}_{\rm ns,1.4}\,{\rm g\,cm^2}$ \citep{2005ApJ...629..979L} with $M_{\rm ns}=1.4\, M_\odot\, M_{\rm ns,1.4}$ the NS mass.} and $\Omega=2\pi/P$ is the angular frequency, given as a function of the spin period. The merger of compact objects usually leaves a fraction of material gravitationally bound to the remnant NS, which subsequently forms an accretion disk and falls back over extended timescales \citep{2008MNRAS.385.1455M, 2017PhRvD..95f3016C, 2017RPPh...80i6901B}. Given an accreting mass ($M_{\rm fb}$), we can define a fallback accretion rate considering one characteristic fall-back timescale ($t_{\rm fb}$) as \citep{2018ApJ...857...95M}
\bary\label{Mdot}
\dot{M} \simeq  \frac23\frac{M_{\rm fb}}{t_{\rm fb}}\left(1 +  \frac{t}{t_{\rm fb}} \right)^{-\frac{5}{3}}\,, 
%\begin{cases}\label{dotM}
%1 \hspace{1.6cm}  t \ll t_{\rm fb}  \,, \cr
%\left( \frac{t}{t_{\rm fb}} \right)^{-\frac{5}{3}}     \hspace{0.4cm}  t_{\rm fb} \ll t   \,, \cr
%\end{cases}
\eary
or two characteristic timescales as
\bary\label{Mdot_2}
\dot{M} \simeq  \frac23\frac{M_{\rm fb}}{t_{\rm fb,e}} 
\begin{cases}\label{dotM}
\left(1 +  \frac{t}{t_{\rm fb, e}} \right)^{-\frac{5}{3}}  \hspace{2.15cm}  t < t_{\rm k}  \,, \cr
\left(1 +  \frac{t_{\rm k}}{t_{\rm fb, e}} \right)^{-\frac{5}{3}}\,\left(1 + \frac{t}{t_{\rm fb,l}} \right)^{-\frac{5}{3}}     \hspace{0.4cm}  t_{\rm k} \leq t   \,, \cr
\end{cases}
\eary

where $t_{\rm k}$ is the temporal accretion break,  $t_{\rm fb, e}$ and $t_{\rm fb, l}$ early and late characteristic fall-back timescale, respectively.  The term $\left(1 +  \frac{t_{\rm k}}{t_{\rm fb, e}} \right)^{-\frac{5}{3}}$ in the second PL segment is introduced so that $\dot{M}$ is smooth around $t_{\rm k}$. Following the merger, a fraction of the ejecta may remain gravitationally bound and subsequently fall back onto the newly formed magnetar.  The interaction between fallback accretion and the magnetar magnetic field defines several characteristic radii that determine the accretion and spin-down regimes of the system. They are the  Alfv\'en ($r_{\rm m}\simeq 22.3 \,{\rm km}\, M^{-\frac17}_{\rm ns, 1.4}\,\dot{M}^{-\frac27}_{-2}\,B^{\frac47}_{15}\,R^{\frac{12}{7}}_{\rm ns,6.1}$), the co-rotation ($r_{\rm c}\simeq 17.4 \,{\rm km}\,  M^{\frac13}_{\rm ns, 1.4} \,P^{\frac{2}{3}}_{-3}$) and the light cylinder ($r_{\rm lc}\simeq 48.5\,{\rm km}\, P_{-3}$) radii,  with $R_{\rm ns}\simeq 1.2\times 10^6\, {\rm cm}\, R_{\rm ns, 6.1}$ the NS radius and $B$ the strength of the dipole magnetic field.  The spin evolution of the magnetar is governed by the competition between magnetic spin-down torques and angular-momentum transfer from fallback accretion. This coupling naturally links the accretion history with the temporal evolution of the spin-down luminosity.  The spin evolution of the magnetar is governed by \citep{2011ApJ...736..108P}
\be\label{dif_eq}
I\frac{d\Omega}{dt}=-N_{\rm dip}+N_{\rm acc}\,.
\ee
The terms $N_{\rm dip}$ and $N_{\rm acc}$ are the spin-down torques from the dipole emission and accretion, respectively.  For $r_{\rm m}\gtrsim R_{\rm ns}$, these torques are $N_{\rm dip} \simeq \frac{\mu^2\Omega^3}{c^3} \frac{r^2_{\rm lc}}{r^2_{\rm m}}$ for $r_{\rm m} \lesssim  r_{\rm lc}$ and $\frac{\mu^2\Omega^3}{c^3}$ for $r_{\rm lc} \lesssim r_{\rm m}$ \citep{2016ApJ...822...33P},  and $N_{\rm acc}=\dot{M}(G\,M_{\rm ns}\,r_{\rm m})^\frac12\, \left[ 1-\left( \frac{r_{\rm m}}{r_{\rm c}} \right)^\frac32 \right]$, respectively, where $\mu=BR_{\rm ns}^3$ is the dipole magnetic moment,  $G$ is the gravitational constant and $c$ the speed of light. The magnetar will accumulate material based on the position of the Alfvén radius in relation to the co-rotation radius. For $r_{\rm m} \lesssim r_{\rm c}$, the magnetar will undergo accretion; otherwise, the system may transition into the propeller regime \cite[e.g., see][]{1998A&ARv...8..279C}.   The spin period that separates both regimes can be calculated, which also corresponds to the steady-state evolution of the system, by requiring the condition $r_{\rm m}= r_{\rm c}$.  In this scenario, the equilibrium spin period is established as $P_{\rm eq}\simeq 1.5\times 10^{-3}\,  {\rm s}\,\, B^{\frac67}_{15}\, R^{\frac{18}{7}}_{\rm ns,6.1}\, M^{-\frac57}_{\rm ns,1.4}\, \dot{M}^{-\frac37}_{-2}$, which is achieved within the time duration specified as $I\Omega_{\rm eq}/\dot{M}(G\,M_{\rm ns}\,r_{\rm m})^\frac12$ with $\Omega_{\rm eq}=2\pi/P_{\rm eq}$. The electromagnetic spin-down luminosity can be calculated by solving

 \be\label{L_sd}
 L_{\rm sd}=\Omega(N_{\rm dip}-N_{\rm acc})\,.
\ee
Different accretion regimes lead to distinct temporal evolutions of the spin-down luminosity.  For $r_{\rm c}\ll r_{\rm m}$,  the equation (\ref{dif_eq}) becomes $\frac{d\Omega}{dt}+\left( \frac{\mu^2}{c^3 I r^2_{\rm m}}+\frac{\dot{M}\,r^2_{\rm m}}{I} \right) \Omega = 0$, which has a solution $\Omega=\Omega_0\, \exp(-\frac{t}{2t_{\rm sd}})$ and therefore the spin-down luminosity becomes
\be\label{Lsd_in}
L_{\rm sd, in}\simeq \left( \frac{\mu^2}{c\, r^2_{\rm m}}+ \dot{M}\,r^2_{\rm m} \right) \Omega^2_0 \exp(-\frac{t}{t_{\rm sd}})\,,
\ee
where $t_{\rm sd}=\frac12 \left( \frac{\mu^2}{c^3 I r^2_{\rm m}}+\frac{\dot{M}\,r^2_{\rm m}}{I} \right)^{-1}$. When $r_{\rm c} = r_{\rm m}$, two solutions can be considered depending on the characteristic timescale.  For $t_{\rm fb}< t$, the accretion rate is approximately constant before the fallback timescale (Eq. \ref{Mdot}), then from  Eqs.  (\ref{dotM}) and (\ref{L_sd})  the spin-down luminosity evolves as

\be\label{Lsd_ea}
L_{\rm sd, bf} =   3.6\times 10^{45}\,{\rm erg\, s^{-1}}\,M^{\frac{17}{7}}_{\rm ns,1.4}\,B^{-\frac{20}{7}}_{15}\,R^{-\frac{60}{7}}_{\rm ns,6.1}\,\left(\frac{t_3}{t_{\rm fb,5}} \right)^{0}\,.
\ee

For a late time ($t\leq t_{\rm fb} $),  the accretion rate evolves as $\dot{M} \propto t^{-\frac{5}{3}}$ and  once the equilibrium is reached \citep{2018ApJ...857...95M},  from equations (\ref{dotM}) and (\ref{L_sd})  the spin-down luminosity  becomes
\be\label{Lsd_late}
L_{\rm sd, af}\simeq1.6\times 10^{42}\,{\rm erg\, s^{-1}}\, B^{-\frac67}_{15}\,M^{\frac{12}{7}}_{\rm ns,1.4}\,R^{-\frac{18}{7}}_{\rm ns,6.1}\,\left(\frac{t_{6}}{t_{\rm fb,5}} \right)^{-\frac{50}{21}}.
\ee

It is important to note that if the evolution of accretion rate were $\dot{M} \propto t^{-\frac{4}{3}}$ as proposed by  \cite{2011MNRAS.413.2031M}, the spin-down luminosity would vary as $L_{\rm sd, af}\propto t^{-\frac{40}{21}}$.   The X-ray flux ($F_{\rm x, sd}$) resulting from internal energy dissipation can be derived from the spin-down luminosity ($L_{\rm sd}$) and the efficiency of converting spin-down energy into radiation $\eta_{\rm x}$. In this case, the X-ray flux can be written as

\bary
F_{\rm x, sd}=K\eta_{\rm x}\, \frac{L_{\rm sd,ll}\,f^{-1}_{\rm b}}{4\pi d_{\rm z}^2\,(1+z)^{\beta_L - 1}}\,,
\eary

where $L_{\rm sd,ll}$ with ${\rm ll = in}$ (Eq. \ref{Lsd_in}),  ${\rm bf}$ (Eq. \ref{Lsd_ea}) or ${\rm af}$  (Eq. \ref{Lsd_late}), $d_z$ is the luminosity distance,  $\beta_{\rm L} \sim 0.8 - 1$, $K$ is a correction to account for the observed XRT band \citep{2001AJ....121.2879B, ber13}, and $f_{\rm b}=1-\cos\theta_j\approx \frac12\theta^2_j$ is the correction factor with $\theta_j$ the half-opening angle,  the essential parameter for defining the cone-shaped region from which the relativistic outflow of a burst emanates.\\

In this framework, the spin-down luminosity is assumed to be partitioned between internal dissipation, which powers the extended X-ray emission, and continuous energy injection into the external shock, which sustains the synchrotron afterglow emission from radio to X-ray frequencies.

\subsubsection{Internal dissipation in a magnetized outflow}

%\subsubsection{Energy dissipation by the Magnetization parameter following the prompt episode}
%

 Following the prompt episode, the spin-down power extracted from the magnetar is assumed to be carried by a relativistic magnetized wind. At early times, the outflow is expected to be Poynting-flux dominated, so magnetic dissipation processes such as magnetic reconnection and internal dissipation can efficiently convert magnetic energy into non-thermal particle acceleration and radiation. In this framework, the magnetization parameter $\sigma$ characterizes the ratio between magnetic and matter energy fluxes 

\be\label{sigma}
\sigma=\frac{L_j}{\dot{M}_j\,c^2}\,,
\ee

and regulates the dissipation efficiency of the outflow. Numerical and analytical studies suggest that magnetic dissipation becomes particularly efficient for moderately to highly magnetized outflows with $\sigma \simeq 10^2 - 3\times 10^3$ \citep[e.g., see][]{2010ApJ...725.2209L, 2012MNRAS.420..483G, 2018ApJ...857...95M}.   The term $L_j=L_{\rm sd,ll}$ represents the spin-down luminosity \citep{2009MNRAS.396.2038B} and $\dot{M}_j$ is the rate at which the baryon loading is ejected from the NS surface. Extended X-ray emission may therefore arise from synchrotron radiation emitted by relativistic particles accelerated within the dissipative magnetized wind.

%Electromagnetic emission in the Poynting-flux-dominated regime is generated by magnetic reconnection, a process that may induce internal shell collisions. In this scenario, the magnetization parameter is a critical factor. Certain magnetic dissipation models predict that the magnetization parameter is comparable to the bulk Lorentz factor and falls within a specific range \citep[$100\lesssim\sigma \lesssim 3000$;][]{2010ApJ...725.2209L, 2012MNRAS.420..483G}.
%The magnetization parameter is defined by
%
%
%\be\label{sigma}
%\sigma=\frac{L_j}{\dot{M}_j\,c^2}\,,
%\ee
%
%where $L_j=L_{\rm sd,ll}$ represents the spin-down luminosity \citep{2009MNRAS.396.2038B} and $\dot{M}_j$ is the rate at which the baryon loading is ejected from the NS surface.

The baryon-loading rate in the magnetized wind can be written as
\be
\dot{M}_j \simeq  \dot{M}_\nu\,f_{\rm cent}
\begin{cases}
%\cases{ 
%0.5 \hspace{0.9cm} r_{\rm m}\lesssim R_{\rm ns} \cr
\frac{R_{\rm ns}}{2\,r_{\rm m}}\hspace{0.8cm}   r_{\rm m} \lesssim  r_{\rm lc}\ \cr
\frac{R_{\rm ns}}{2\,r_{\rm m}} \hspace{0.8cm} r_{\rm lc} \lesssim r_{\rm m}\,, \cr
\end{cases}
\ee
with  $\dot{M}_\nu=\dot{M}_{\rm \nu, ob}(t)+\dot{M}_{\rm \nu, acc}(t)$, $f_{\rm cent}=e^{(\frac{P_{\rm c}}{P})^\frac32}$  and   $P_{\rm c}\simeq 2.7\,R^{\frac12}_{\rm ns}\, r^{-\frac12}_{\rm m}\,M^{-\frac12}_{1.4}$.     The terms $\dot{M}_{\rm \nu, ob}(t)$ and $\dot{M}_{\rm acc}$ are associated with the mass loss rate due to different  sources of neutrinos. The term $\dot{M}_{\rm \nu, ob}(t)$  defined by \citep{2011MNRAS.413.2031M}
\be\label{M_nu-ob}
\dot{M}_{\rm \nu, ob}(t)=3\times10^{-4}\,(1+\frac{t}{t_{\rm kh}})^{-\frac52}\,e^{-\frac{t}{t_{\rm thin}}}\,M_{\odot}\,s^{-1}\,,
\ee
is due to the neutrino ablation,  and 
\be\label{M_nu-acc}
\dot{M}_{\rm acc}(t)=1.2\times10^{-5}\,M_{\rm 1.4}\,\dot{M}_{-2}^{\frac53}\,M_{\odot}\,s^{-1}\,,
\ee
is  due to the accretion \citep{2011ApJ...736..108P}.   The cooling timescale $t_{\rm kh}\approx 2\,{\rm s}$ corresponds to  Kelvin-Helmholtz  and  $t_{\rm thin}\approx (10 - 30)\,{\rm s}$  to the timescale when NS becomes optically thin to neutrinos \citep{2018ApJ...857...95M}.\\

While the extended X-ray component may arise from internal dissipation powered by the magnetar spin-down luminosity, the radio and optical/IR afterglow are naturally interpreted within the synchrotron forward-shock framework. We therefore describe below the synchrotron evolution before and after the jet-break phase.

\subsubsection{Synchrotron FS light curves before the jet-break phase}

Continuous energy injection from the long-lived magnetar can refresh the external shock and naturally account for the shallow plateau phase observed in the optical/IR bands. The energy injected continuously by the millisecond magnetar ($L_{\rm sd,bf}\propto t^0$) on the afterglow can produce refreshed shocks.  Given the Blanford-McKee solution \citep{1976PhFl...19.1130B}  during the deceleration phase with the isotropic-equivalent kinetic energy $E_{\rm K}= \eta f^{-1}_{\rm b} L_{\rm sd}\,t_{\rm fb}$,  the bulk Lorentz factor evolves as $\Gamma\propto t^{-\frac{1}{4}}$. The factor $\eta$ denotes the fraction of the initial rotational energy transferred to the afterglow. The relation between the isotropic-equivalent kinetic energy and the isotropic gamma-ray energy is given through its efficiency

\be
\eta_\gamma=\frac{E_{\rm \gamma, iso}}{E_{\rm \gamma, iso} + E_{\rm K}}\,.
\ee

The synchrotron spectral breaks and the maximum flux in the uniform-density medium become $\nu_{\rm m}\propto t^{-1}$, $\nu_{\rm c}\propto t^{-1}$ and $F_{\rm max}\propto t$, respectively. The spectral breaks at the self-absorption regime evolve as  $\nu_{\rm a}\propto t^{\frac{1}{5}}$ for $\nu_{\rm a}<\nu_{\rm m}<\nu_{\rm c}$, as $\nu_{\rm a}\propto t^{-\frac{p}{p+4}}$ for $\nu_{\rm m}<\nu_{\rm a}<\nu_{\rm c}$ and as $\nu_{\rm a}\propto t^{\frac{1}{5}}$ for $\nu_{\rm a}<\nu_{\rm c}<\nu_{\rm m}$. 
The synchrotron light curves before the jet-break phase is explicitly written in Appendix \ref{App_A}.  It is worth noting that the spin-down luminosity for the late time is too steep ($L_{\rm sd,af}\propto t^{-\frac{50}{21}}$) to significantly change the synchrotron light curves \citep{2006ApJ...642..354Z}. Therefore, standard synchrotron light curves could be used \citep{1998ApJ...497L..17S}.

\subsubsection{Synchrotron FS light curves after the jet-break phase}

The steep achromatic decay observed after the temporal break is difficult to reproduce within the standard post-jet-break synchrotron scenario assuming constant microphysical parameters. Motivated by this behavior, we explore a phenomenological evolution of $\epsilon_e$ and $\epsilon_B$ during the post-jet-break phase. The jet-break transition occurs when the relativistic beaming angle becomes comparable to the jet opening angle, i.e. when $\Gamma^{-1} \sim \theta_j$. In the standard synchrotron forward-shock scenario, the microphysical parameters $\epsilon_e$ and $\epsilon_B$ are commonly assumed to remain constant throughout the afterglow evolution. However, several observational studies have suggested that temporal evolution of these parameters may occur in some GRBs \citep[e.g., see][]{2003ApJ...597..459Y, 2006MNRAS.369.2059P, 2006MNRAS.370.1946G, 2006A&A...458....7I, 2005PThPh.114.1317I, 2020ApJ...905..112F}. Such variations may arise from changes in particle-acceleration efficiency and magnetic-field amplification as the relativistic shock propagates through the circumburst medium, although the detailed physical mechanism responsible for this evolution remains uncertain.  Nevertheless, phenomenological evolution of the microphysical parameters has previously been shown to alleviate tensions between observed temporal decays and standard synchrotron closure relations.

During the post-jet break phase, the microphysical parameters evolve as 
$\epsilon_{\rm e}\propto t^{-a}$ and $\epsilon_{\rm B}\propto t^{-b}$ and the bulk Lorentz factor as $\Gamma\propto t^{-\frac12}$. Therefore, the spectral breaks and the maximum synchrotron flux scale as  $\nu_{\rm m} \propto t^{-\frac{4(1+a)+b}{2}}$, $\nu_{\rm c} \propto t^{\frac{3b}{2}}$ and $F_{\rm max} \propto t^{-\frac{2+b}{2}}$, respectively. Breaks in the self-absorption regime evolve as  $\nu_{\rm a}\propto t^{-\frac{1-5a+b}{5}}$ for $\nu_{\rm a}<\nu_{\rm m}<\nu_{\rm c}$, or $\nu_{\rm a}\propto t^{-\frac{4(p+1)+4a(p-1)+b(p+2)}{2(p+4)}}$ for $\nu_{\rm m}<\nu_{\rm a}<\nu_{\rm c}$. The synchrotron light curves with microphysical parameter evolution during the post jet-break phase are explicitly written in Appendix \ref{App_B}.  These expressions show that temporal evolution of the microphysical parameters can substantially steepen the post-jet-break synchrotron decay while preserving standard values of the electron spectral index.

\subsection{Description of the multi-wavelength observations}

The multi-wavelength observations of GRB~250704B/EP250704a reveal distinct temporal behaviours across the radio, optical/IR, and X-ray bands. The optical/IR observations exhibit a long-lasting plateau phase followed by a steep achromatic decay, while the X-ray emission shows evidence for both extended internal activity and afterglow radiation. The radio observations provide additional constraints on the synchrotron spectral evolution and the post-jet-break behaviour of the external shock. In the following discussion, we interpret these features within a framework involving synchrotron FS emission with continuous energy injection together with internally dissipated spin-down emission from a long-lived magnetar.

All optical/IR fluxes exhibited a plateau phase followed by a steep achromatic decay. The spectral indices measured at $2.7\times10^{3}$ and $1.5\times10^{5}$ s after the trigger remain nearly unchanged ($\beta_{\rm o}=0.72 \pm 0.01$ and $0.64 \pm 0.08$, respectively; see Fig.~\ref{fig3:x-ray_and_optical}), suggesting that the temporal break at $\approx7.7\times10^{4}$ s is achromatic and therefore can be interpreted as a jet-break transition. The steep decay observed after the temporal break is naturally explained by the synchrotron evolution described by Eqs.~(\ref{slow_mic_v1}) and (\ref{slow_mic_v2}). In particular, the optical/IR observations can be reproduced under the spectral regime $\nu_m < \nu_{\rm o,II} < \nu_c$ or $\nu_a < \nu_{\rm o,II} < \nu_c$ for an electron spectral index of $p = 2.6 \pm 0.2$.

Within this framework, the temporal evolution of the microphysical parameters is constrained through the relation
\[
\alpha_{\rm o,II} - p =
\left(
\frac{p+1}{4}
\right)b + (p-1)a.
\]
It is important to note that the observed optical/IR decay could only be reproduced within the standard post-jet-break synchrotron scenario (assuming constant microphysical parameters $a=0$ and $b=0$) unless an unusually large value of $p \approx 3.4$ were adopted. Therefore, the observed steep achromatic decay favours a scenario involving temporal evolution of the microphysical parameters during the post-jet-break phase.

Before the jet break, the plateau phase observed across all optical/IR filters is is reproduced through synchrotron FS scenario with continuous energy injection from a long-lived millisecond magnetar, evolving as $F_\nu \propto t^{(3-p)/2}$ for the cooling conditions $\nu_m < \nu_{\rm o,I} < \nu_c$ or $\nu_a < \nu_{\rm o,I} < \nu_c$ (Eqs. \ref{slow1_before_inj} or \ref{slow2_before_inj}).

The radio upper limits provide additional constraints on the synchrotron spectral evolution during the post-jet-break phase. In particular, the radio observations favour the synchrotron evolution described by Eq.~(\ref{slow_mic_v1}) with $\nu_a < \nu_R < \nu_m$ or $\nu_R < \nu_a$, since the alternative evolution described by Eq.~(\ref{slow_mic_v2}) would require an unusually steep value of $b < -4$ to reproduce the radio observations \citep{2003MNRAS.339..881R, 2005PThPh.114.1317I, 2006MNRAS.369..197F, 2006MNRAS.369.2059P, 2013arXiv1305.3689L,2013MNRAS.428..845L, 2018ApJ...859..163H, 2020ApJ...905..112F, 2024MNRAS.527.1884F, 2026MNRAS.545f1970F}.

The X-ray observations are naturally explained by a superposition of internal dissipation powered by the magnetar spin-down luminosity and synchrotron FS emission. Before the jet break, the spectral index ($\beta_x = 0.72 \pm 0.01$) derived from the SED at $2.7 \times 10^{3}$ s indicates that the optical/IR and X-ray fluxes evolve within the same spectral segment, $\nu_m < \nu_{\rm o} < \nu_x < \nu_c$, for $p = 2.6 \pm 0.2$.\footnote{It is worth highlighting that the {\itshape Swift}/XRT team reported a value of $\Gamma_{\rm x}=1.93^{+1.02}_{-0.49}$ at $7.7\times 10^3\,{\rm s}$, which could be consistent with the X-ray evolution in the PL segment $ \nu_{\rm x}<\nu_{\rm c}$ or $ \nu_{\rm c}<\nu_{\rm x}$.} Under these conditions, the synchrotron forward-shock component evolves as $F_\nu \propto t^{(2-p)/2}$ in the X-ray band.  At later times, the synchrotron cooling break likely crossed the XRT band, while the internally dissipated spin-down component likely provided a comparable contribution to the observed X-ray flux.

\subsection{Constraints on the physical parameters}

To constrain the physical parameters of the magnetar-powered energy-injection scenario, we performed a joint fit to the X-ray, optical/IR, and radio observations of GRB~250704B/EP250704a. The modelling simultaneously accounts for the internally dissipated spin-down emission and the synchrotron FS scenario, including the continuos energy injection and temporal evolution of the microphysical parameters before and after the jet-break phase, respectively.  For the internal energy dissipation of the magnetar spin-down power, we consider the fallback accretion rate with one (Eq. \ref{Mdot}) and two characteristic fall-back timescales (Eq. \ref{Mdot_2}) for value of fallback mass of $M_{\rm fb}=0.8\,M_\odot$ \citep{2018ApJ...857...95M} and efficiency $\eta_{\rm x}=0.02$ \citep{2013ApJ...775...67B, 2019ApJ...878...62X}. When we consider a fallback accretion rate with a single characteristic timescale, we fix the characteristic fall-back time to $t_{\rm fb}=7.7\times 10^4\,{\rm s}$, which corresponds to the temporal breaks found in each filter, and   when we consider two characteristic timescales, we fix them with $t_{\rm fb, e}=7.1\times 10^2\,{\rm s}$ and $t_{\rm fb,l}=7.7\times 10^4\,{\rm s}$ for $t_{\rm k}=3.1\times 10^3\,{\rm s}$.  We normalize the synchrotron FS light curves at $h\nu_{\rm x}=1\,{\rm keV}$ and $\nu_{\rm R}=10\, {\rm GHz}$ for the X-ray\footnote{The flux density measured at 10 keV is converted to its equivalent value at 1 keV using the PL index and the appropriate conversion factor
reported in \cite{2010A&A...519A.102E}} and radio observations, respectively, and at $\nu_{\rm o_k}=2.398$, $3.356$, $4.008$, $4.862$ and  $6.397\times 10^{15}\,{\rm Hz}$ for the ${\rm k=J}$-, ${\rm z}$-, ${\rm i}$-, ${\rm r}$- and ${\rm g}$-optical bands, respectively.  The luminosity distance is estimated assuming a spatially flat $\Lambda$CDM cosmology with $H_0 = 67.4$ km s$^{-1}$ Mpc$^{-1}$ and $\Omega_M = 0.315$ \citep{2020A&A...641A...6P}.

The analytical model described in Section \ref{model} depends on the parameter set
$\Sigma_{\sigma_\rho}=\{E_{\rm K}, n, p, \varepsilon_{\rm B}, \varepsilon_{\rm e}, \zeta, a, b, B, P \}$, where $\zeta$ represents the fraction of electrons accelerated into the non-thermal population.   The adopted prior ranges were chosen to encompass physically plausible values commonly inferred for sGRBs and magnetar-powered afterglows. In particular, the isotropic-equivalent kinetic energy and circumburst density were allowed to vary within the ranges typically reported for sGRBs, while the microphysical parameters were restricted to values below equipartition. The electron spectral index was constrained to the interval expected from relativistic shock acceleration theory. Similarly, the initial spin period and dipolar magnetic-field strength were selected within the range commonly inferred for millisecond magnetars formed after compact-object mergers. These choices were intended to avoid imposing strong prior-driven constraints while ensuring physically meaningful solutions. The adopted priors are weakly informative, physically motivated, and restricted to ensure meaningful shock solutions with $\varepsilon_{\rm e} + \varepsilon_{\rm B} < 1$. The electron spectral index was restricted to values expected from relativistic particle acceleration at collisionless shocks. %The parameter $\zeta$ corresponds to the fraction of electrons accelerated into the non-thermal population at the shock front.

A likelihood function assuming normally distributed residuals was adopted. Posterior distributions were sampled using the No-U-Turn Sampler (NUTS) implemented within the PyMC3 framework \citep{peerj-cs.55}. Posterior predictive realizations were generated to evaluate the consistency of the inferred parameter distributions with the observed multi-wavelength evolution.

   The best-fit values of the isotropic-equivalent kinetic energy ($E_{\rm K}$), circumburst density ($n$), the spectral PL index ($p$), the microphysical parameters ($\varepsilon_{\rm B}$ and $\varepsilon_{\rm e}$), the fraction of electrons accelerated by the shock front ($\zeta$),  PL indexes (${\rm a}$ and ${\rm b}$), initial spin period (P), and dipolar magnetic-field (B) obtained from the MCMC analysis are listed in Table \ref{tab5:fit}.  Figure \ref{fig:mcmc} presents a corner plot that illustrates the 1-dimensional marginalized distributions that follow for each parameter, when a timescale of fallback accretion is considered.  The corner plots corresponding to the two-timescale scenario are not shown because the inferred parameter distributions are similar to those presented in Figure \ref{fig:mcmc}. It is important to note that a Bayesian method for assessing a model's capacity to explain observed data is through posterior predictive checks \citep{2017arXiv170901449G}.
   
The posterior distributions converge toward a parameter region capable of reproducing the observed multi-wavelength evolution. The resulting best-fit values are listed in Table \ref{tab5:fit} and are used in the following section to discuss the physical implications of the model.

\subsection{Physical interpretation of the multi-wavelength modelling}\label{sec:results}

\begin{figure}
    \centering
    \includegraphics[width=0.48\linewidth]{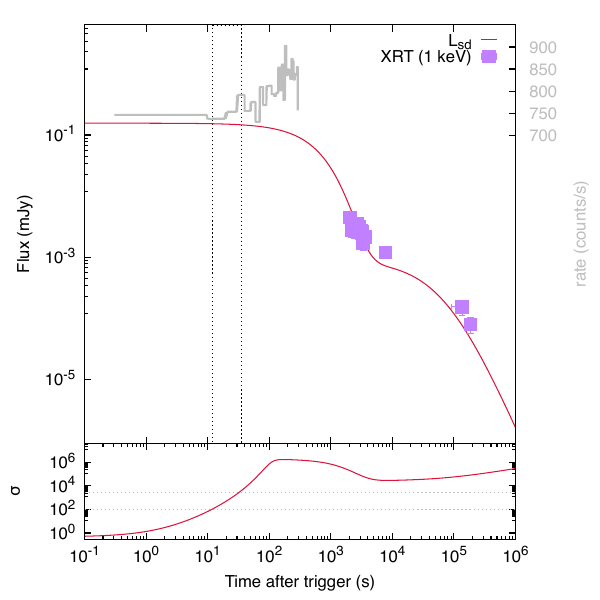}
    \includegraphics[width=0.48\linewidth]{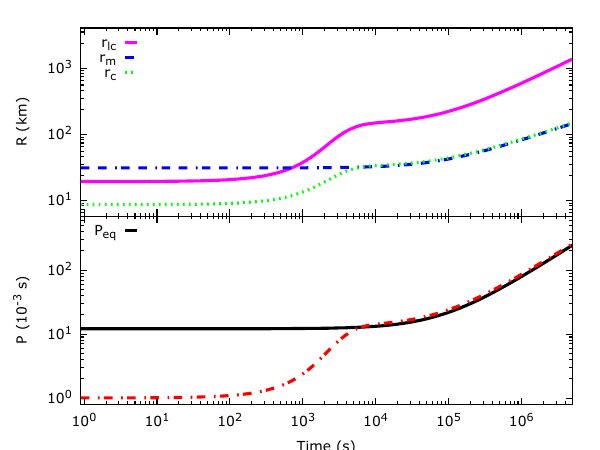}
    \includegraphics[width=0.85\linewidth]{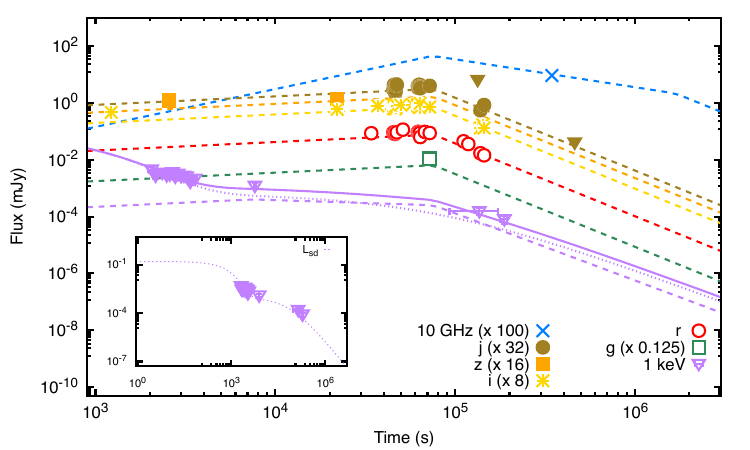}    
    \caption{Best-fit temporal evolution obtained assuming a fallback accretion rate with a single characteristic timescale. Upper-left panel: X-ray observations together with the internally dissipated spin-down component (red curve) and the temporal evolution of the magnetization parameter (lower sub-panel). The gray curve shows the HXM2 (CALET) light curve in the 7–10 keV band. Dashed vertical lines indicate the interval during which the magnetization parameter evolves within the range $10^2 \lesssim \sigma \lesssim 3\times10^3$. Upper-right panel: temporal evolution of the light-cylinder, co-rotation, and Alfvén radii together with the spin-period evolution. Lower panel: multi-wavelength observations with the best-fit synchrotron forward-shock and internally dissipated spin-down components. Dashed lines correspond to the synchrotron forward-shock contribution, dotted lines to the internally dissipated spin-down component, and solid black lines to the total emission.}
    \label{fig:multi_fit_v1}
\end{figure}

\begin{figure}
    \centering
    \includegraphics[width=1\linewidth]{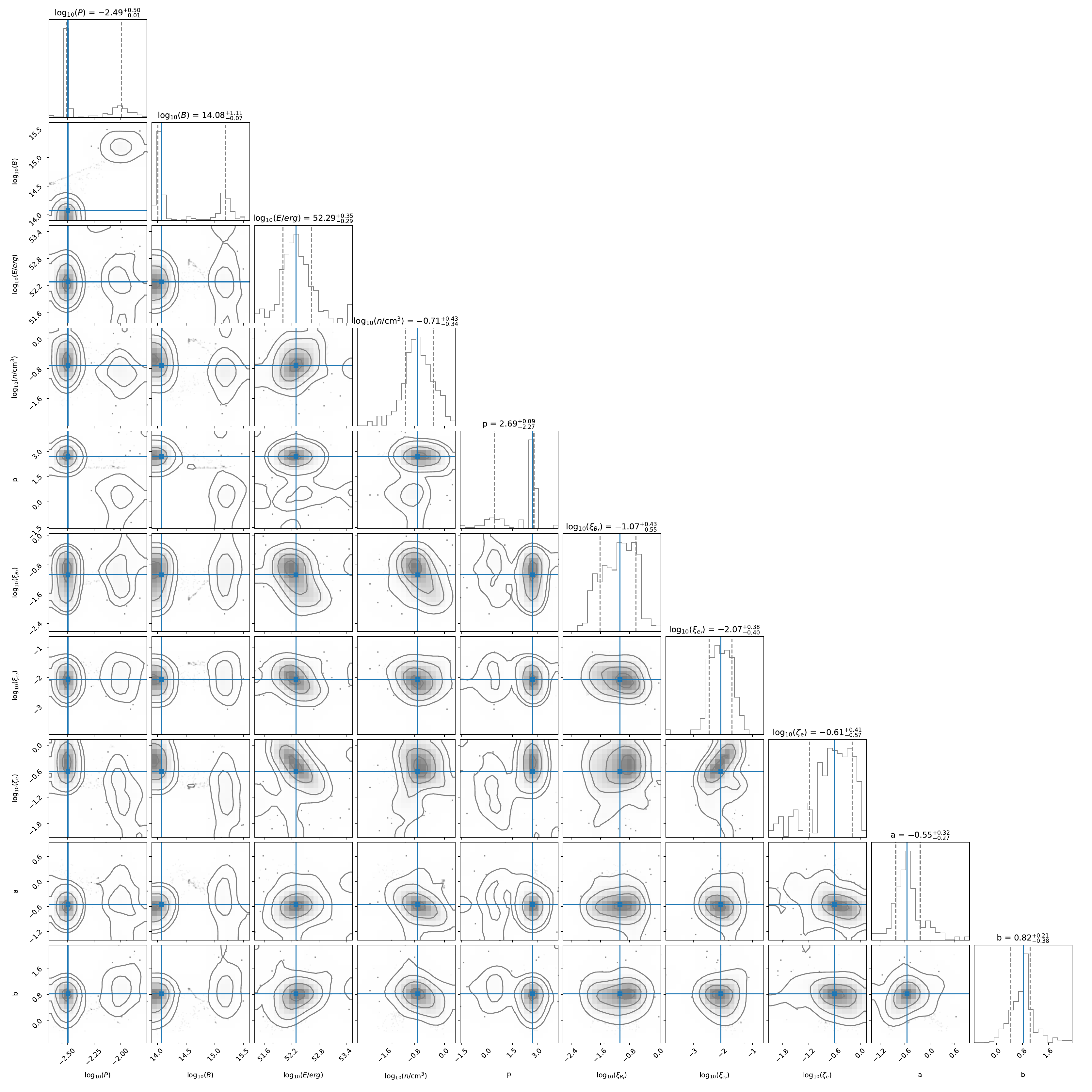}
    \caption{Corner plot showing the posterior distributions obtained from the MCMC analysis assuming a fallback accretion rate with a single characteristic timescale. The contours correspond to the $1\sigma$, $2\sigma$, and $3\sigma$ confidence regions.}
    \label{fig:mcmc}
\end{figure}

\begin{figure}
    \centering
    \includegraphics[width=0.48\linewidth]{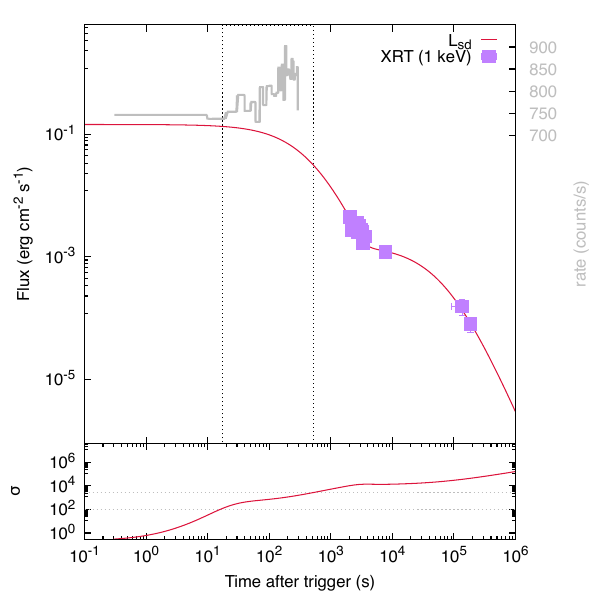}
    \includegraphics[width=0.48\linewidth]{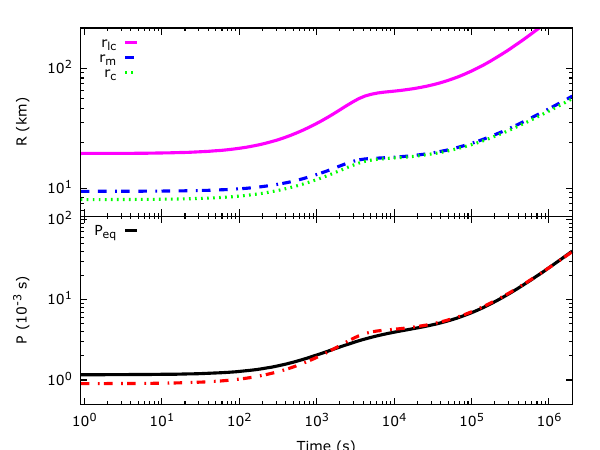}
    \includegraphics[width=0.85\linewidth]{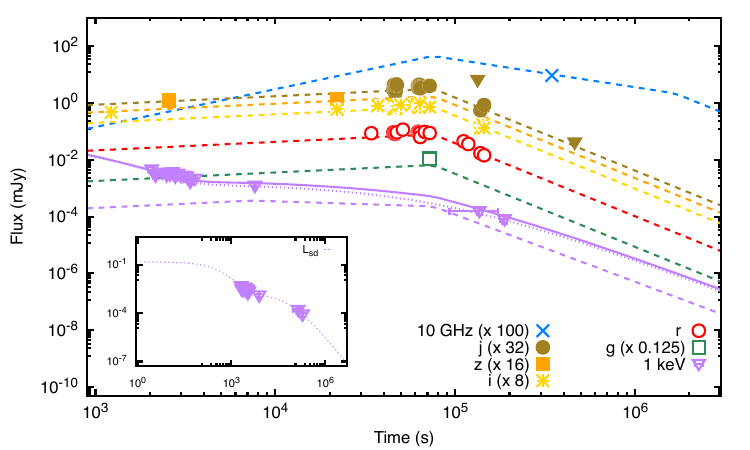}    
    \caption{Same as Figure \ref{fig:multi_fit_v1}, but assuming a fallback accretion rate characterized by two distinct timescales.}
    \label{fig:multi_fit_v2}
\end{figure}

The best-fit solutions obtained from the joint multi-wavelength modelling support a long-lived magnetar central engine powering both the extended X-ray activity and the refreshed synchrotron afterglow emission. Figures \ref{fig:multi_fit_v1} and \ref{fig:multi_fit_v2} show the resulting temporal evolution obtained assuming fallback accretion with one and two characteristic timescales, respectively.

Figure~\ref{fig:multi_fit_v1} shows the best-fit multi-wavelength evolution (upper) obtained assuming a fallback accretion rate with a single characteristic timescale. Although the internally dissipated spin-down component reproduces the extended X-ray emission, an additional synchrotron FS contribution is required at late times, particularly beyond $\sim10^5$ s. Furthermore, the upper sub-panel shows the HXM2 (CALET) light curve (gray) on counts in the 7 - 10 keV channel. 

The magnetization parameter evolves within the range $10^2 \lesssim \sigma \lesssim 3\times10^3$ during the interval associated with the extended X-ray activity, corresponding to the regime where magnetic dissipation is expected to become efficient in the relativistic outflow.  The evolution of the characteristic radii indicates that the system initially evolves in the propeller regime ($r_c < r_m$), while the spin period gradually approaches the equilibrium configuration at $\sim10^4$ s. The joint modelling successfully reproduces the observed multi-wavelength evolution through the combination of internally dissipated spin-down emission and synchrotron FS radiation with continuous energy injection.

Figure~\ref{fig:multi_fit_v2} shows analogous results assuming a fallback accretion rate characterized by two distinct timescales. The main results are i) a better description of the X-ray data, although a minor contribution from a synchrotron FS is necessary, and ii) the evolution of the magnetization parameter in the range where magnetic dissipation processes are expected to become efficient and the duration of the HXM2 (CALET) light curve, as shown in the left-hand panel.  The magnetization parameter evolves in the range of $10^2 \leq \sigma\leq 3\times10^3$ during the interval [12.2: 520.4] s.

The temporal coincidence between the onset of the extended emission and the evolution of the magnetization parameter toward the regime where magnetic dissipation may become efficient suggests a possible connection between both processes.   A similar evolution of the characteristic radii and spin period is obtained in the two-timescale scenario. In this scenario, the spin period reaches equilibrium at $\sim 10^3\,{\rm s}$, then deviates slightly, and finally reaches $\sim 10^4\,{\rm s}$. A minor contribution from the synchrotron FS scenario is observed in the X-ray data, as shown in the lower panel of Figure~\ref{fig:multi_fit_v2}.

Based on the best-fit values, we calculate relevant quantities and present a discussion of the results as outlined below. Because both fallback-accretion scenarios yield nearly identical physical parameters (Table \ref{tab5:fit}), we focus the discussion on the one-timescale solution for clarity.\\

\subsubsection{The circumburst density, bulk Lorentz factor and energetic}

The circumburst density provides important constraints on the progenitor environment through the properties of the parsec-scale surrounding medium. The best-fit density, $n\approx 0.15\,{\rm cm^{-3}}$, is consistent with the low-density environments typically inferred for sGRBs, supporting their compact-object merger origin and their significant offsets from star-forming regions \citep{2006ApJ...650..261S, 2014ARA&A..52...43B, 2013ApJ...776...18F, 2022ApJ...940...56F, 2022MNRAS.515.4890O}. For the sGRB population, the circumburst density spans the range $10^{-3}\lesssim n \lesssim1\,{\rm cm^{-3}}$, with a median value of $n\lesssim0.15\,{\rm cm^{-3}}$ \citep{2014ARA&A..52...43B, 2015ApJ...815..102F, 2020MNRAS.495.4782O}.

The isotropic-equivalent kinetic energy, $E_{\rm K}\approx 1.2\times10^{52}\,{\rm erg}$, lies among the highest values reported for optically detected sGRBs \citep{2014ARA&A..52...43B, Fong_2015}. Combining the isotropic-equivalent kinetic and gamma-ray energies, $E_{\rm \gamma,iso}\approx 4\times10^{51}\,{\rm erg}$, we obtain a gamma-ray efficiency of $\eta_\gamma\approx0.25$. This value is consistent with the mean efficiency $<\eta_\gamma>=0.40^{+0.49}_{-0.35}$ reported by \citet{Fong_2015} for a sample of sGRBs modeled with $\epsilon_{\rm B}=0.01$. Using the inferred isotropic-equivalent kinetic energy and spin-down luminosity, $L_{\rm sd}\simeq 10^{47}\,{\rm erg\,s^{-1}}$, the fraction of the initial rotational energy transferred to the observed afterglow is $\eta\approx1.3\times 10^{-2}$, consistent with values inferred in other magnetar-powered scenarios \citep{2013ApJ...775...67B, 2019ApJ...878...62X}.

The temporal break at $t_{\rm j}=7.7\times10^{4},{\rm s}$ observed in the multi-wavelength data is interpreted as the jet-break transition. The inferred circumburst density and isotropic-equivalent kinetic energy imply an initial bulk Lorentz factor of $\Gamma\approx60$ at $2\times10^3\,{\rm s}$ and a jet opening angle of $\theta_{\rm j}\approx3^\circ$. These derived values are consistent with the distribution of jet opening angles and the typical lower limits ($\gtrsim3^\circ$) inferred for sGRBs \citep{2014ARA&A..52...43B, Fong_2015, 2019MNRAS.489.2104T, 2023ApJ...959...13R}.

\subsubsection{Evolution of microphysical parameters}

The steep achromatic decay observed after the jet break suggests temporal evolution of the microphysical parameters during the late afterglow phase. Such evolution has previously been invoked to explain plateau phases, evolving spectral breaks, and tensions in multi-wavelength afterglow modelling \citep{2003MNRAS.339..881R, 2003ApJ...597..459Y, 2005PThPh.114.1317I, 2006MNRAS.369.2059P, 2006MNRAS.370.1946G, 2006A&A...458....7I, 2006MNRAS.369..197F, 2020ApJ...905..112F}.

During the post-jet-break phase, the microphysical parameters evolve as $\epsilon_{\rm e}\propto t^{-a}$ and $\epsilon_{\rm B}\propto t^{-b}$, with $a\approx -0.5$ and $b\approx 0.9$. These values imply an increase in $\epsilon_{\rm e}$ and a decrease in $\epsilon_{\rm B}$ during the late afterglow evolution.   Using the inferred parameter values, the synchrotron spectral breaks evolve as $\nu_{\rm a}\propto t^{-0.9}$ for $\nu_{\rm a}<\nu_{\rm m}<\nu_{\rm c}$ or $\nu_{\rm a}\propto t^{-1.2}$ for $\nu_{\rm m}<\nu_{\rm a}<\nu_{\rm c}$, while $\nu_{\rm m}\propto t^{-1.5}$, $\nu_{\rm c}\propto t^{1.5}$, and the maximum flux density evolves as $F_{\rm max}\propto t^{-1.5}$. Consequently, the synchrotron light curves evolve as $F_\nu\propto t^{0.5}$ for $\nu<\nu_{\rm a}$, $F_\nu\propto t^{-1}$ for $\nu_{\rm a}<\nu<\nu_{\rm m}$, $F_\nu\propto t^{-2.6}$ for $\nu_{\rm m}<\nu<\nu_{\rm c}$, and $F_\nu\propto t^{-1.8}$ for $\nu_{\rm c}<\nu$. These temporal scalings successfully reproduce the steep achromatic evolution observed during the post-jet-break phase.

\subsubsection{Prompt Episode}

Photospheric, shock-breakout, and cocoon scenarios generally predict a significant thermal component during the prompt emission phase \citep{2006ApJ...652..482P, 2009ApJ...702.1211R, 2012ApJ...747...88N}. The time-integrated KW spectrum of GRB~250704B/EP250704a is well described by a Band function characterized by $E_{\rm p}=970\pm100.6\,{\rm keV}$, $\alpha_{\gamma}=-1.160\pm0.086$, and $\beta_{\gamma}=-2.28\pm1.13$.

The absence of a clear thermal component during the prompt episode suggests that the relativistic outflow may be significantly magnetized, carrying a substantial fraction of its energy in magnetic form. Alternatively, the observed Band spectrum may arise from reprocessed quasi-thermal emission through dissipation of kinetic \citep{2009ApJ...700L..65Z, 2015ApJ...801..103G} or magnetic energy close to the photosphere \citep{2006ApJ...652..482P, 2010ApJ...725.1137L, 2016ApJ...831..175V, 2015MNRAS.454L..31A, 2013MNRAS.428.2430L, 2011ApJ...738...77V, 2012ApJ...761L..18V}.  The non-thermal spectral shape together with the spectral evolution therefore support a scenario involving a moderately to highly magnetized, Poynting-flux-dominated outflow \citep{2009ApJ...700L..65Z, 2015ApJ...801..103G, 2014MNRAS.445.3892B}.

\subsubsection{Extended X-ray component emission}

The detection by HETE-2\footnote{The High Energy Transient Explorer-2} of a soft long-tail component following the initial hard spike of GRB~050709 \citep{2005Natur.437..855V} revealed an additional emission component that has since been identified in only a limited number of sGRBs \citep[e.g., GRB~050724, 061006, 061210, 080503, 090715, 090916, 100625A, 211211A, 230307A and others;][]{2005Natur.437..855V, 2009ApJ...696.1871P, 2014ARA&A..52...43B, 2015ApJ...811....4K, 2019ApJ...877..147K, 2022Natur.612..228T, 2023ApJ...943..146C, 2025NSRev..12E.401S}.

Several physical mechanisms have been proposed to explain this extended X-ray emission, including the onset of the early afterglow \citep{2009ApJ...696.1871P}, jet-cocoon interactions \citep{10.1093/mnras/stx2357}, cocoon emission \citep{2024PASJ...76..550O}, spin-down activity from long-lived millisecond magnetars \citep{2008MNRAS.385.1455M, 2018ApJ...857...95M, 2019ApJ...877..147K}, and powerful winds launched from accretion disks \citep{2009ApJ...699L..93L}. Nevertheless, the physical origin of this component remains debated.

The onset of the extended X-ray emission closely coincides with the magnetization parameter entering the regime where magnetic dissipation is expected to operate efficiently, both in the single-timescale (Fig.~\ref{fig:multi_fit_v1}) and two-timescale (Fig.~\ref{fig:multi_fit_v2}) fallback-accretion scenarios. In contrast, the termination of the extended emission approximately occurs when the magnetization parameter leaves this efficient dissipation regime in the two-timescale scenario. These results strongly suggest that the extended X-ray emission is powered by internal dissipation of the magnetar spin-down luminosity regulated by the outflow magnetization.

\begin{figure}
    \centering
    \includegraphics[width=0.85\linewidth]{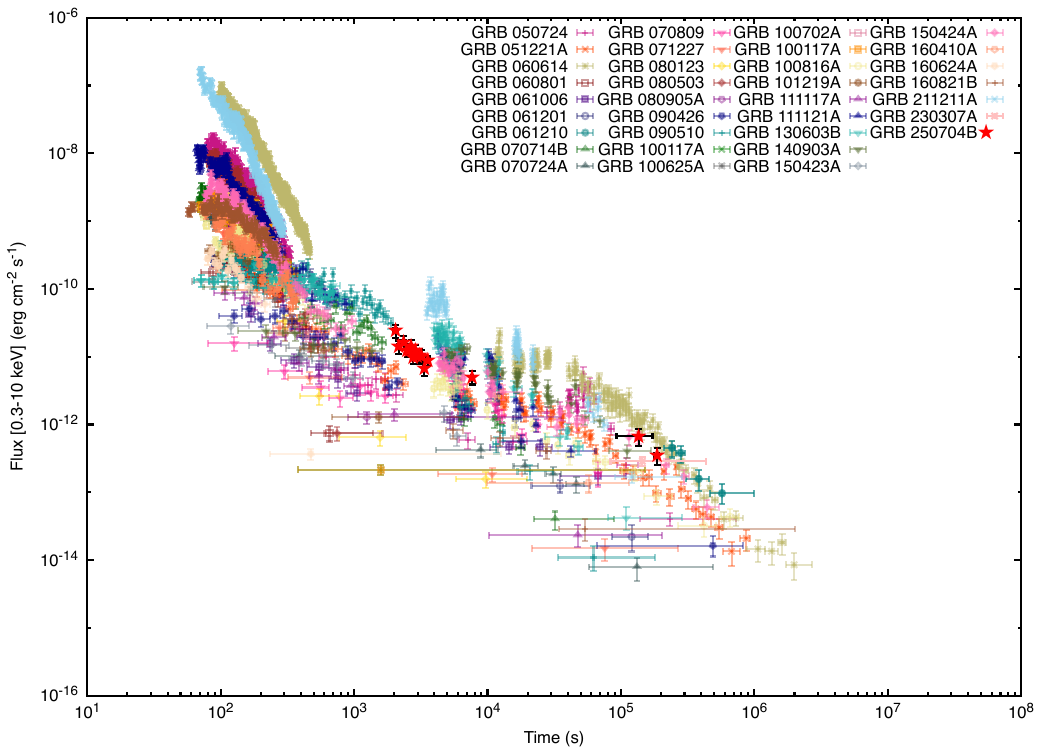}    
    \caption{X-ray light curves of sGRBs exhibiting extended emission detected by Swift/XRT, including GRB~250704B/EP250704a (star symbol). The light curves are shown in the observer frame.}
    \label{fig:berger}
\end{figure}

In this framework, fallback accretion naturally prolongs the duration of the extended-emission phase, allowing a more sustained release of spin-down power and providing a better description of the observed X-ray evolution \citep{2018ApJ...857...95M}. Similar conclusions were reached by \citet{2019ApJ...877..147K}, who analyzed a sample of 26 sGRBs and found that long-lived magnetars provide a plausible explanation for the extended X-ray emission observed in these events.  Figure~\ref{fig:berger} compares the 0.3–10 keV Swift/XRT light curves of GRB~250704B/EP250704a with those of other sGRBs exhibiting extended emission. This comparison places GRB~250704B/EP250704a among the brightest late-time extended-emission sGRBs observed to date.

\subsubsection{The derived features of the magnetar scenario}

The magnetar scenario has been extensively investigated as a mechanism for powering long-lasting activity in GRBs through spin-down energy release and continuous energy injection 
\citep{1998MNRAS.298...87D, 1998A&A...333L..87D, 2001ApJ...552L..35Z, 2006MNRAS.372L..19F, Troja2007, Lyons2010, ber13, 2013MNRAS.430.1061R, lu14, 2018ApJ...857...95M, li18}. In particular, \citet{ber13}, based on a sample of eight GRBs with measured redshifts detected by \textit{Swift}/BAT between 2005 and 2009, showed that late-time activity, including plateau phases and precursor emission, is difficult to reconcile with standard accretion-driven models unless the central engine is a newly formed magnetar. Building upon this framework, \citet{lu14} analyzed a larger sample of 214 GRBs with known redshifts and classified them into four categories (Gold, Silver, Aluminium, and others) according to the likelihood of magnetar-powered emission. More recently, \citet{li18} applied a comparable classification scheme to 101 GRBs using \textit{Swift}/XRT light curves.

More generally, plateau phases observed in GRB afterglows—characterized by a shallow decay in X-rays and, in some cases, optical bands followed by a rapid steepening—are difficult to explain within the standard external-shock scenario assuming prompt BH formation. Instead, they point toward sustained energy injection from a long-lived central engine. These features are naturally explained by the formation of a rapidly rotating, highly magnetized NS, for which the plateau phase traces the gradual release of spin-down energy, as suggested for events such as GRB~180620A \citep[e.g.,][]{Becerra2019,Zou2022} and GRB~190829A \citep[e.g.,][]{2021ApJ...918...12F}.

For GRB~250704B/EP250704a, the best-fit values of dipolar magnetic-field strength and initial spin period are $B\approx 1.3\times10^{14}\,{\rm G}$, and $P\approx3.2\,{\rm ms}$, respectively.  These values place GRB~250704B/EP250704a within the Silver category defined by \citet{lu14}, and are fully consistent with the properties inferred for other sGRB magnetar candidates. In particular, the dipolar magnetic-field strengths, spin periods, and spin-down luminosities reported by \citet{lu14} span the ranges $(0.37\pm0.01)\leq B\leq(11.57\pm4.21)\times10^{15}\,{\rm G}$, and $(0.96\pm0.13)\leq P\leq(5.55\pm1.25)\,{\rm ms}$, respectively.

\subsubsection{Origin of the long-lasting optical/IR plateau}

Plateau phases were first identified in X-rays and later confirmed at optical wavelengths. The earliest evidence of an optical plateau was reported for GRB~050801 \citep{Vestrand2006}, which showed a pronounced flattening in its light curve at optical wavelengths. This behaviour suggested that prolonged energy injection plays an important role in shaping the afterglow evolution \citep{Rykoff2006,Vestrand2006}. These observations suggested that plateau phases are closely connected to prolonged central-engine activity and sustained energy injection.

Building on these observational developments, \cite{2022ApJS..261...25D} collected the optical afterglow data of 179 GRBs with known redshifts that exhibited a plateau phase.  Optical observations obtained with different filters by {\itshape Swift}/UVOT, RATIR, and the Subaru Telescope were rescaled to the r-band. The time at the end of the plateau emission reported for the seven sGRBs; GRB~000301C, 050922C, 060313A, 080913A, 081228A, 090510A and 190627A, was $\approx 7.59\times 10^5$, $5.89\times 10^3$, $3.78\times 10^3$, $3.47\times 10^5$, $6.32\times 10^2$, $4.18\times 10^3$ and $5.37\times 10^4\,{\rm s}$, respectively, indicating that the optical plateau of GRB 250704B/EP250704a is long-lasting but not unprecedented among sGRBs.

Based on the derived parameters, the spin-down timescale is shorter than the characteristic fallback timescale. Therefore, the duration of the optical/IR plateau is primarily governed by the long-lasting fallback accretion onto the magnetar remnant. In this scenario, fallback accretion continuously feeds the central engine and sustains prolonged energy injection into the forward shock, naturally producing the extended optical/IR plateau observed in GRB 250704B/EP250704a. These results further support the presence of a long-lived magnetar undergoing sustained fallback accretion after the compact-object merger.

Long-lived plateau emission is not unique to GRB 250704B. Similar behavior has also been reported in the short-duration GRBs GRB 150423A and GRB 220831A, both of which exhibited extended X-ray plateau phases lasting up to $\sim10^{4}$--$10^{5}$ s despite being classified as sGRB/sGRB-EE events \citep{2017A&A...607A..84K, 2025MNRAS.537.2061F}. These bursts further support the existence of a subclass of merger-driven GRBs powered by long-lived central engines, most plausibly millisecond magnetars, although the detailed physical origin of the plateau may differ from event to event.

\subsubsection{Comparison with other merger-magnetar candidates}

Despite their markedly different prompt emission properties, GRB 250704B/EP250704a exhibits several similarities to the merger-driven magnetar candidates GRB 211211A and GRB 230307A at later times. These include prolonged central-engine activity and evidence for extended high-energy emission. GRB 250704B exhibited a canonical short-duration prompt episode followed by extended soft X-ray activity and a long-lasting optical/IR plateau ending in an abrupt achromatic decay. By contrast, GRB 211211A and GRB 230307A displayed unusually long prompt emission episodes accompanied by strong evidence for luminous kilonova emission \citep[e.g., see][]{2022Natur.612..228T,2022Natur.612..223R, 2024Natur.626..737L}.

The most plausible interpretation for GRB~211211A and GRB~230307A involves a compact binary merger producing a long-lived magnetar remnant instead of an immediately formed BH \citep{2022Natur.612..228T, 2023ApJ...943..146C, 2023ApJ...954L..29D, 2024MNRAS.529L..67D, 2024ApJ...963L..26Z, 2025NSRev..12E.401S, 2026MNRAS.549ag949F}. Additional similarities emerge from studies of the jet composition and magnetic evolution. For instance, \citet{2024MNRAS.529L..67D} analyzed the jet composition in GRB~211211A and concluded that the outflow is dominated by Poynting flux based on the absence of detectable thermal emission. Similarly, \citet{2023ApJ...943..146C} investigated the thermal and non-thermal emission of GRB~211211A within a hybrid jet scenario and inferred a significant magnetized component in the outflow. In the case of GRB~230307A, \citet{2024ApJ...963L..26Z} interpreted the early X-ray plateau followed by a sharp decline observed in a soft X-ray band (0.3  - 4 keV)  as evidence for a magnetar wind undergoing gradual magnetic dissipation. These authors argued that the presence of an X-ray plateau strongly supports a magnetar central engine with evolving radiation efficiency.

The magnetic evolution of the relativistic jet in GRB~211211A and GRB~230307A was further discussed by \citet{2024MNRAS.529L..67D} and \citet{2024ApJ...963L..26Z}, respectively. The similarities between the prompt-emission properties of GRB~211211A and GRB~230307A were already noted in their discovery papers \citep[e.g., see][]{2022Natur.612..228T,2022Natur.612..223R, 2024Natur.626..737L, 2025NSRev..12E.401S}. Subsequently, \citet{2024ApJ...969...26P} carried out a more detailed comparative study, showing that both bursts exhibit very short minimum variability timescales and consistent photospheric-emission properties.

%In addition, \citet{2024ApJ...969...26P} performed a comparative analysis of GRB~211211A and GRB~230307A, showing that both bursts share several remarkable similarities, including very short minimum variability timescales and consistency between their photospheric-emission properties.

In contrast, the multi-wavelength evolution of GRB~250704B/EP250704a can be self-consistently interpreted through the combined contribution of internally dissipated spin-down emission and synchrotron forward-shock radiation with continuous energy injection and evolving microphysical parameters. Unlike GRB~230307A, no pronounced X-ray dip or clear accretion-to-propeller transition is required in our modelling. These differences likely reflect diversity in the fallback-accretion history, magnetization evolution, and viewing geometry among merger-powered magnetar remnants. Similar to GRB~211211A and GRB~230307A, the observational properties of GRB~250704B/EP250704a favour a moderately to highly magnetized outflow, consistent with a Poynting-flux-dominated jet.Therefore, similarly to GRB~211211A and GRB~230307A, the multi-wavelength properties of GRB~250704B/EP250704a are naturally explained by a magnetized relativistic outflow in which magnetic energy dissipation plays a key role in powering the observed emission.

\subsubsection{Comparison with previous interpretations}

GRB~250704B/EP250704a exhibited extended X-ray emission together with a day-long optical plateau followed by an abrupt late-time decay, features that are difficult to reconcile within the standard synchrotron afterglow framework. An alternative interpretation based on an off-axis structured jet has recently been proposed by \citet{2025arXiv250902769S} using afterglowpy. However, reproducing such a long-lasting plateau requires an unusually large Lorentz factor ($\Gamma\leq1000$) in the jet core, and the off-axis interpretation remains uncertain because most GRBs are generally believed to be observed close to the jet axis \citep{2015ApJ...799....3R, 2023ApJ...958..126F, 2024MNRAS.533.1629O}. Consequently, this scenario is not adopted in the present work.

On the other hand, \citet{2026arXiv260114137L} reported a minutes-long ($\sim560$ s) episode of soft X-ray emission immediately following GRB~250704B. The detection of this extended soft X-ray component by \textit{Einstein Probe} supports the presence of prolonged central-engine activity in an event otherwise resembling a canonical short GRB. These authors further suggested that extended soft X-ray emission may represent a common feature of merger-driven bursts.

In this work, we present an alternative interpretation in which the multi-wavelength evolution of GRB~250704B/EP250704a is powered by a long-lived accreting millisecond magnetar. In our scenario, the extended X-ray emission arises from internal dissipation within a relativistic magnetized outflow powered by the magnetar spin-down luminosity, while the optical/IR and radio emission are explained through synchrotron forward-shock radiation with continuous energy injection. The steep achromatic decay is naturally reproduced by allowing temporal evolution of the microphysical parameters during the post-jet-break phase. Within this framework, the X-ray observations can be interpreted as a superposition of internally dissipated spin-down emission and synchrotron forward-shock radiation.

Although geometric effects may contribute to the observed evolution, they are not strictly required to explain the multi-wavelength behaviour of GRB~250704B/EP250704a. Instead, a magnetar-powered scenario involving both internal dissipation and refreshed synchrotron emission provides a self-consistent interpretation of the temporal and spectral evolution of the burst.

\subsubsection{Alternative scenarios can reproduce some observational properties, but fail to simultaneously explain the full multi-wavelength evolution}

Based on the temporal and spectral analyses, we briefly examine alternative scenarios that may account for some features of the observed multi-wavelength evolution. In particular, we explore: (i) a synchrotron scenario in the thick-shell regime, (ii) synchrotron evolution with early-time variation of the microphysical parameters, and (iii) a synchrotron scenario involving stratified ejecta.

\paragraph{Synchrotron FS light curves during the thick-shell regime.}
During the thick-shell regime,  the bulk Lorentz factor evolves as $\Gamma\propto t^{-\frac14}$.  The
spectral breaks and the maximum flux of synchrotron emission evolving in the constant-density environment are  $\nu_{\rm a}\propto t^{\frac15}$, $\nu_{\rm m}\propto t^{-1}\,$, $\nu_{\rm c}\propto t^{-1}$ and $F_{\rm max}\propto t\,$, respectively. The respective synchrotron light curves evolve as $F_{\rm \nu} \propto t^{\frac12}  \nu^{-\frac12}$ for $\nu_{\rm c}<\nu<\nu_{\rm m}$ and $\propto t^{\frac{2-p}{2}} \nu^{-\frac{p}{2}}$ for $\nu_{\rm m}<\nu$ (fast-cooling regime), and $F_{\rm \nu} \propto t^{-\frac{p-3}{2}} \, \nu^{-\frac{p-1}{2}}$ for $\nu_{\rm m}<\nu<\nu_{\rm c}$ and $\propto t^{-\frac{p-2}{2}} \nu^{-\frac{p}{2}}$ for $\nu_{\rm c}<\nu$ \citep[slow-cooling regime;][]{2013NewAR..57..141G, 2013ApJ...776..120Y}. The temporal and spectral indexes of optical/IR fluxes ($\alpha_{\rm o,I}\approx -0.13$ and $\beta_{o}=0.72\pm0.01$ at $2700\,{\rm s}$) are consistent with the evolution of synchrotron light curves $F_{\rm \nu} \propto t^{-\frac{p-3}{2}} \, \nu^{-\frac{p-1}{2}}$ for $\nu_{\rm m}<\nu_{\rm o,I} <\nu_{\rm c}$ with $p=2.6\pm 0.2$. Taking into account the spectral index $\beta_{\rm x}=0.72\pm0.01$  derived from the SED at $2\times 10^3\,{\rm s}$, the X-rays would evolve first as $F_{\rm \nu}\propto t^{\frac{3-p}{2}}$ for $\nu_{\rm m}<\nu_{\rm x}<\nu_{\rm c}$ and then as $\propto t^{\frac{2-p}{2}}$ for  $ \nu_{\rm c} < \nu_{\rm x}$. In this scenario, additional X-ray component would be necessary because  X-ray observations are not consistent with evolution  $\propto t^{\frac{3-p}{2}}$ with $p=2.6\pm 0.2$. On the other hand, the deceleration time at $2\times 10^3\,{\rm s}$ is extremely larger than the duration of the burst \citep[$T_{\rm 90}=0.68^{+0.16}_{-0.14}\,{\rm s}$;][]{2025GCN.40940....1S} which  is naturally explained by the dynamics of the thin-shell regime. 

\paragraph{Synchrotron FS light curves with early evolution of  microphysical parameters.}

Given the lack of understanding of the processes behind the energy transfer from protons to electrons and magnetic fields in relativistic shocks, it is reasonable to assume the evolution of microphysical parameters during the afterglow phase.  During the deceleration phase, the bulk Lorentz factor in the uniform-density medium evolves as $\Gamma\propto t^{-\frac{3}{8}}$.  The synchrotron spectral breaks and the maximum flux in the uniform-density medium vary as $\nu_{\rm m}\propto t^{-\frac{3+4a+b}{2}}$, $\nu_{\rm c}\propto t^{\frac{3b-1}{2}}\,$ and $F_{\rm max}\propto t^{-\frac{b}{2}}\,$, respectively. The predicted synchrotron light curves in the uniform-density medium are $F_{\rm \nu} \propto t^{\frac{b-1}{4}} \nu^{-\frac12}$ for $\nu_{\rm c}<\nu<\nu_{\rm m}$, and 
$\propto t^{\frac{2(1+2a+b) - p(3 + 4a + b)}{4}} \nu^{-\frac{p}{2}}$ for $\nu_{\rm m}<\nu$ (fast-cooling regime), and $F_{\rm \nu} \propto t^{\frac{3+4a-b - p(3 + 4a + b)}{4}} \nu^{-\frac{p-1}{2}}$ for $\nu_{\rm m}<\nu<\nu_{\rm c}$ and  $\propto t^{\frac{2(1+2a+b) - p(3 + 4a + b)}{4}} \nu^{-\frac{p}{2}}$ for $\nu_{\rm c}<\nu$ \citep[slow-cooling regime;][]{2024MNRAS.527.1884F}.  For PL indexes $a=0$ and $b=0$, the standard synchrotron light curves in a constant-density medium is recovered \citep{1998ApJ...497L..17S}.  Considering that X-ray and optical fluxes evolving in the same PL segment $\nu_{\rm m}<\nu<\nu_{\rm c}$ at $2.7\times 10^3$ s, and  with $p=2.6\pm 0.2$ and $\alpha_{\rm o,I}\approx -1.3$, the values of PL indexes are constrained by equation $-1.48=1.78a + b$. Similarly, considering $\alpha_{\rm x,I}\approx 0.65$ and the fact that the synchrotron cooling break must pass through the XRT band at some point before $\approx 10^{5}\,{\rm s}$, the values of PL indexes are additionally constrained by the equation $-6=10.67a + b$ and $3b -1 <0$. Given the spectral index $\beta_{\rm x}=0.72\pm0.01$ at $2700\,{\rm s}$, the X-rays would evolve as $F_{\rm \nu}\propto t^{\frac{3-p}{2}}$ for $\nu_{\rm m}<\nu_{\rm x}<\nu_{\rm c}$ which is inconsistent with the observations and therefore an additional emission of the X-ray component would be necessary to describe these observations.  In this complex scenario, the microphysical parameters would evolve twice (before and after the break at $7.7\times 10^{4}\,{\rm s}$). This behaviour has not been observed or reported, which posing challenges for theoretical models.

\paragraph{Synchrotron FS light curves with stratified ejecta.}  Instantaneous energy injection by central engine activity is associated with an ejected mass moving with a wide range of Lorentz factors, usually described by a PL distribution. The ejected mass can acquire a velocity structure $M(>\Gamma) \propto \Gamma^{-\alpha_s}$, where the front of the ejected material is faster than the back \citep{2001ApJ...551..946T}. The bulk Lorentz factor in the constant-density environment varies as $\Gamma\propto t^{-\frac{3}{\alpha_s+7}} $ and the synchrotron flux as $F_{\rm \nu} \propto t^{-\frac{3(2p-\alpha_s-1)}{7+\alpha_s}}$ for $\nu_{\rm m}<\nu<\nu_{\rm c}$ and  $F_{\rm \nu} \propto t^{-\frac{2(3p-\alpha_s-1)}{7+\alpha_s}}$  for $\nu_{\rm c} < \nu$ \citep[e.g, see][]{2000ApJ...535L..33S, 2006ApJ...642..354Z, 2015MNRAS.448..417B, 2019ApJ...871..200F}.  Taking into account that X-ray and optical fluxes evolve in the same PL segment $\nu_{\rm m}<\nu_{\rm o} <\nu_{\rm x} <\nu_{\rm c}$ with $p=2.6$, the value of the PL index of the velocity structure would be $\alpha_{\rm s}\approx 4.07$.  It is worth noting that the relation of the indexes $q$ and $\alpha_{\rm s}$ would be $q=\frac{10-2\alpha_{\rm s}}{7+\alpha_{\rm s}}\approx 0.17$ which is different from that expected for the magnetar scenario. Although the s-value favours the BH as progenitor, long plateaus explained in this scenario with energy injection would require unreasonable parameters \citep{2025arXiv250902769S}.

Overall, the temporal and spectral evolution of GRB~250704B/EP250704a can be consistently interpreted within a long-lived magnetar framework involving internal dissipation powered by spin-down luminosity together with synchrotron FS emission sustained by continuous energy injection and microphysical parameter evolution.

\section{Conclusions}
\label{sec:summary}

GRB~250704B/EP250704a adds to the growing population of short GRBs, including recent events detected by \textit{Einstein Probe}, that challenge the simplest merger-driven afterglow scenarios. The optical/IR observations revealed a long-lasting plateau phase followed by a rapid achromatic decay. Together with the X-ray evolution, these features can be naturally interpreted within a long-lived magnetar central-engine scenario involving continuous energy injection into the circumburst medium.

In this framework, the extended X-ray emission following the prompt episode is attributed to internal dissipation within a relativistic magnetized outflow powered by the spin-down luminosity of a millisecond magnetar. A better description of the extended X-ray evolution is obtained when the fallback accretion rate is modeled using two characteristic timescales.

Using the best-fit parameters derived from the joint modelling of the X-ray, optical/IR, and radio observations, we find that before the jet break the optical/IR emission is naturally explained by synchrotron FS radiation evolving between the characteristic and cooling spectral breaks in a uniform-density medium. The X-ray observations are reproduced through a superposition of internally dissipated spin-down emission and synchrotron forward-shock radiation, initially evolving within the same spectral regime before the synchrotron cooling break crosses the XRT band.

The steep late-time decay observed in the optical/IR bands, together with the radio and X-ray evolution at $\gtrsim10^5\,{\rm s}$, can be reproduced by allowing temporal evolution of the microphysical parameters during the post-jet-break phase. Within this framework, the observed temporal evolution can be directly linked to the intrinsic physical properties of the newly formed compact object, without requiring extreme geometric configurations or finely tuned viewing angles.

The inferred circumburst density is low, consistent with typical sGRB environments, while the relatively small jet opening angle favours the observability of the achromatic post-break evolution. Overall, our results support a scenario in which at least a fraction of merger-driven sGRBs are powered by long-lived magnetar remnants capable of sustaining prolonged multi-wavelength activity through continued fallback accretion and spin-down energy release.

Overall, our results support a scenario in which at least a fraction of merger-driven sGRBs are powered by long-lived magnetar remnants capable of sustaining prolonged multi-wavelength activity through continued fallback accretion and spin-down energy release. The inferred magnetization evolution together with the absence of a prominent thermal component further supporting a moderately to highly magnetized, Poynting-flux-dominated outflow. Future polarization measurements and broadband monitoring of similar events will help constrain the magnetic composition and dissipation properties of merger-powered relativistic outflows.

\section*{Acknowledgements}

NF  acknowledges  financial  support  from UNAM-DGAPA-PAPIIT  through  grant  IN112525. BBK is supported by IBS under the project code IBS-R018-D3. M.G.D. acknowledges the support of the JSPS Grant-in-Aid for Scientific Research (KAKENHI) (A), Grant Number JP25H00675.  AG was supported by Universidad Nacional Autónoma de México Postdoctoral Program (POSDOC).

%%%%%%%%%%%%%%%%%%%%%%%%%%%%%%%%%%%%%%%%%%%%%%%%%%
\section*{Data Availability}

No new data were generated or analysed in support of this research.



\bibliographystyle{mnras}
\bibliography{RS_microp_mnras}
% if your file is called example.bib

\newpage
\clearpage

% \usepackage{multirow}
% \usepackage{graphicx}
\begin{table}
\centering
\caption{Best-fit values from the spectral analysis of the {\itshape KW} data in the 18 to 1160 keV energy range, obtained using the Band function.}
\label{tab1:fit}
%\resizebox{0.3\columnwidth}{!}{%
\begin{tabular}{lcc}
\hline
\hline
Parameter & Value & $\chi^{2}/{\rm ndf}$               \\ \hline \hline

%\multicolumn{3}{l}{$\gamma$-rays (Spectra)} [18-1160~keV]              \\ \hline
$E_{\rm p}$  (keV)      & $970\pm100.6$     & \multirow{3}{*}{$0.08/1$} \\
$\alpha_\gamma$                  & $-1.16\pm0.086$   &                    \\
$\beta_\gamma$                   & $-2.28\pm1.13$    &                    \\ \hline

\end{tabular}%
%}
\end{table}

\begin{table}
\centering
\caption{The best-fit values for the HXM2 ({\itshape CALET}/CGBM) light curves across five channels were determined using linear fitting.}
\label{tab2:fit}
%\resizebox{0.3\columnwidth}{!}{%
\begin{tabular}{lcc}
\hline
\hline
Slope & Value & Band
\\ \hline\hline %\multicolumn{3}{l}{X-rays (counts)}               \\ \hline 
$m_{\rm 1}$         & $0.41\pm0.06$     &  [7 - 10~keV] \\
$m_{\rm 2}$         & $0.18\pm0.06$     &  [10 - 25~keV]   \\
$m_{\rm 3}$         & $0.05\pm0.05$     &  [25 - 50~keV]   \\
$m_{\rm 4}$         & $0.04\pm0.04$     &  [50 - 100~keV]   \\
$m_{\rm 5}$         & $0.01\pm0.04$     &  [100 - 170~keV]   \\

\hline

\end{tabular}%
\\Note: The entire data are obtained from the {\itshape CALET}/ website.\footnote{\url{https://cgbm.calet.jp/cgbm\_trigger/flight/1435651911/index.html}} 
\end{table}

\begin{table}
\centering
\caption{The best-fit values of the count rate in the 0.3-10~keV band considering the 2-break fit. The spectral indexes $\beta_{\rm x, 1}$ and $\beta_{\rm x, 2}$ are reported for $3.2\times 10^3$ and $7.7\times 10^3\,{\rm s}$, respectively. The spectral index $\beta_{\rm x}$ is obtained together with optical observations at $2.7\times 10^3\,{\rm s}$.} 
\label{tab3:fit}
%\resizebox{0.3\columnwidth}{!}{%
\begin{tabular}{lcc}
\hline
\hline
Parameter & Value & $\chi^{2}/{\rm ndf}$               \\ \hline \hline
%\multicolumn{3}{l}{X-rays (Light curve)} [0.3-10~keV]              \\ \hline 
$t_{\rm b_x,1}$ (s)        & $(3.4^{+36.9}_{-1.3})\times10^{3}$     & \multirow{5}{*}{$5.5/10$}  \\
$t_{\rm b_x,2}$ (s)        & $(1.2^{+0.9}_{-1.2})\times10^{5}$     &                    \\
$\alpha_{\rm x, I}$       & $1.4^{+3.6}_{-0.6}$     &                         \\
$\alpha_{\rm x, II}$       & $0.65^{+0.18}_{-0.61}$     &                         \\
$\alpha_{\rm x, III}$       & $2.1^{+5.9}_{-1.9}$     &                         \\\hline
$\beta_{\rm x, 1}$       & $0.96^{+0.16}_{-0.13}$     &                 \\
$\beta_{\rm x, 2}$       & $0.93^{+1.02}_{-0.49}$     &                 \\
\hline
$\beta_{\rm x}$       & $0.72\pm 0.01$     &                 \\

\hline\hline

\end{tabular}%
%}
\\Note: These values are reported by the \textit{Swift}/XRT Team at the \textit{Swift} page.\footnote{\url{https://www.swift.ac.uk/xrt\_live_cat/00019908/}} 
\end{table}

\begin{table}
\centering
\caption{The best-fit values of Optical/IR afterglow observations with  SPL and BPL functions. The spectral indexes are $\beta_{\rm o} = 0.72 \pm 0.01$ and $0.64 \pm 0.01$ for $2.7\times 10^3\, {\rm s}$ and $1.5\times 10^5\, {\rm s}$, respectively.}
\label{tab4:fit}
%\resizebox{0.3\columnwidth}{!}{%
\begin{tabular}{lcccc}
\hline
\hline
Band & $\alpha_{o, I}$  & $t_{\rm b_o}\,(\times10^{5}\,{\rm s} )$ & $\alpha_{o, II}$ & $\chi^{2}/{\rm ndf}$               \\ \hline \hline

%\multicolumn{3}{l}{[r band]}               \\ \hline 
j  & $-0.13$     & $0.76\pm0.73$ &  $3.35$  & 4.7/10\\
z  & $-0.13$     & $-$ &  $-$  & 2.3/4\\               
i  & $-0.12$     & $0.78\pm0.68$ &  $3.30$ & 5.4/10\\
r  & $-0.10$     & $0.75\pm0.71$ &  $3.29$  & 4.3/10\\
g         & $-$     & $-$ &  $-$  & $-$\\                 

\\\hline

\end{tabular}%
%}
\end{table}

\begin{table}
\centering
\renewcommand{\arraystretch}{1.4}
\setlength{\tabcolsep}{24pt} %
\caption{Median values of the parameters required for modelling GRB 250704B. These values are derived with a fallback accretion rate with one and two characteristic timescales, and are determined by applying symmetrical quantiles.}
\label{tab5:fit}
%\resizebox{0.3\columnwidth}{!}{%
\begin{tabular}{lccc}
\hline
\hline
Parameter & Prior & Values  & Values              \\ 
          &       & (one timescale)       & (two timescales)\\\hline \hline
%\multicolumn{3}{l}{$\gamma$-rays (Spectra)} [18-1160~keV]              \\ \hline
%{\bf Synchrotron scenario} & &\\
${\rm log_{10}[E_{\rm K}\,({\rm erg})]}$ & $U(50,54)$ & $52.28_{-0.29}^{+0.34}$  & $52.26_{-0.05}^{+0.04}$   \\
${\rm log_{10}[n\,(cm^{-3})}]$    & $U(-3,0)$  & $-0.71_{-0.33}^{+0.42}$ & $-0.83_{-0.05}^{+0.02}$ \\
${\rm p}$                 & $U(2.01,3.5)$ & $2.69_{-2.27}^{+0.08}$  & $2.40_{-0.01}^{+0.01}$ \\
${\rm log_{10} [\varepsilon_{\rm B}]}$    & $U(-2,-0.5)$  & $-1.06_{-0.54}^{+0.43}$ & $-1.03_{-0.04}^{+0.02}$  \\
${\rm log_{10} [\varepsilon_{\rm e}]}$    & $U(-2,-1)$  & $-2.06_{-0.39}^{+0.38}$ & $-2.12_{-0.06}^{+0.09}$  \\
${\rm log_{10} [\zeta_{\rm e}]}$    &  $U(-3,0)$ & $-0.60_{-0.57}^{+0.40}$ & $-0.68_{-0.06}^{+0.08}$  \\
${\rm a}$                 & $U(-1,1)$ & $-0.55_{-0.27}^{+0.31}$  & $-0.47_{-0.02}^{+0.02}$ \\
${\rm b}$                &  $U(-1,1)$ & $0.82_{-0.38}^{+0.21}$  & $0.88_{-0.02}^{+0.01}$  \\
${\rm  log_{10} [B\, (G)]}$   &  $U(13.5,15.5)$  & $14.07\substack{+1.11 \\ -0.06}$  &$14.33\substack{+1.11 \\ -0.06}$\\
${\rm log_{10} [P\,(s)]}$            &  $U(-5,-1)$  & $-2.49\substack{+0.50 \\ -0.01}$  & $-2.68\substack{+0.53 \\ -0.02}$\\

\hline
\end{tabular}%
%}
\end{table}

\newpage
\clearpage

% Checar si es necesario incluir las References: \cite{2014ARA&A..52...43B, 2019ApJ...877..147K}

\appendix

\section{Synchrotron FS light curves before the jet-break phase}\label{App_A}

The predicted synchrotron light curves before the jet-break phase in the uniform-density medium are \citep{2006ApJ...642..354Z, 2013NewAR..57..141G}
{\small
\begin{eqnarray}
\label{fast_before_inj}
F_{\rm \nu} \propto
\begin{cases} 
t  \nu^{2},\hspace{2cm} \nu<\nu_{\rm a}, \cr
t^{\frac{4}{3}}  \nu^{\frac{1}{3}},\hspace{1.68cm} \nu_{\rm a}<\nu<\nu_{\rm c}, \cr
t^{\frac{1}{2}}  \nu^{-\frac{1}{2}},\hspace{1.5cm} \nu_{\rm c}<\nu<\nu_{\rm m}, \cr
t^{\frac{2-p}{2}}  \nu^{-\frac{p}{2}},\hspace{1.1cm} \nu_{\rm m}<\nu, 
\end{cases}
\end{eqnarray}
}

for the fast-cooling regime, and
{\small
\begin{eqnarray}
\label{slow1_before_inj}
F_{\rm \nu} \propto
\begin{cases} 
t  \nu^{2},\hspace{1.76cm} \nu<\nu_{\rm a}, \cr
t^{\frac{4}{3}}  \nu^{\frac{1}{3}},\hspace{1.5cm} \nu_{\rm a}<\nu<\nu_{\rm m}, \cr
t^{\frac{3-p}{2}}  \nu^{-\frac{p-1}{2}},\hspace{0.7cm} \nu_{\rm m}<\nu<\nu_{\rm c}, \cr
t^{\frac{2-p}{2}}  \nu^{-\frac{p}{2}},\hspace{1cm} \nu_{\rm c}<\nu
\end{cases}
\end{eqnarray}
}

or

{\small
\begin{eqnarray}
\label{slow2_before_inj}
F_{\rm \nu} \propto
\begin{cases} 
t  \nu^{2},\hspace{1.72cm} \nu<\nu_{\rm m}, \cr
t^{\frac{3}{2}}  \nu^{\frac{5}{2}},\hspace{1.5cm} \nu_{\rm m}<\nu<\nu_{\rm a}, \cr
t^{\frac{3-p}{2}}  \nu^{-\frac{p-1}{2}},\hspace{0.7cm} \nu_{\rm a}<\nu<\nu_{\rm c}, \cr
t^{\frac{2-p}{2}}  \nu^{-\frac{p}{2}},\hspace{0.9cm} \nu_{\rm c}<\nu, 
\end{cases}
\end{eqnarray}
}

for the slow-cooling regime.

\section{B. Synchrotron FS light curves with microphysical parameter evolution during the post jet-break phase.} \label{App_B}

The synchrotron light curves with microphysical parameter evolution during the post jet-break phase are given by

{\small
\begin{eqnarray}
\label{slow_mic_v1}
F_{\rm \nu} \propto
\begin{cases} 
t^{-a}  \nu^{2},\hspace{2.64cm} \nu<\nu_{\rm a}, \cr
t^{-\frac{1-2a+b}{3}}  \nu^{\frac{1}{3}},\hspace{2cm} \nu_{\rm a}<\nu<\nu_{\rm m}, \cr
t^{\frac{4a-b-p(4+4a+b)}{4}}  \nu^{-\frac{p-1}{2}},\hspace{0.7cm} \nu_{\rm m}<\nu<\nu_{\rm c}, \cr
t^{\frac{2(2a+b)-p(4+4a+b)}{4}}  \nu^{-\frac{p}{2}},\hspace{0.7cm} \nu_{\rm c}<\nu\,,
\end{cases}
\end{eqnarray}
}

for $\nu_{\rm a}<\nu_{\rm m}<\nu_{\rm c}$ and 

{\small
\begin{eqnarray}
\label{slow_mic_v2}
F_{\rm \nu} \propto
\begin{cases} 
t^{-a}  \nu^{2},\hspace{2.6cm} \nu<\nu_{\rm m}, \cr
t^{\frac{4+b}{4}}  \nu^{\frac{5}{2}},\hspace{2.48cm} \nu_{\rm m}<\nu<\nu_{\rm a}, \cr
t^{\frac{4a-b-p(4+4a+b)}{4}}  \nu^{-\frac{p-1}{2}},\hspace{0.7cm} \nu_{\rm a}<\nu<\nu_{\rm c}, \cr
t^{\frac{2(2a+b)-p(4+4a+b)}{4}}  \nu^{-\frac{p}{2}},\hspace{0.7cm} \nu_{\rm c}<\nu\,, 
\end{cases}
\end{eqnarray}
}

for $\nu_{\rm m}<\nu_{\rm a}<\nu_{\rm c}$.

%%%%%%%%%%%%%%%%%%%%%%%%%%%%%%%%%%%%%%%%%%%%%%%%%%

% Don't change these lines
\bsp	% typesetting comment
\label{lastpage}
\end{document}